\documentclass[useAMS,usenatbib]{mn2e}
\usepackage{graphicx} 
\usepackage{longtable}
\usepackage{hyperref}
\usepackage{amsmath}
\usepackage{lscape}
\usepackage{color}
\newcommand{\mnras}{MNRAS}

\newcommand{\aap}{A\&A}

\newcommand{\apj}{ApJ}
\newcommand{\apjl}{ApJL}
\newcommand{\aj}{AJ}

\newcommand{\apjs}{ApJS}
\newcommand{\apss}{Ap\&SS}

\newcommand{\actaa}{Acta Astron.}
\newcommand{\procspie}{Proc. SPIE}

\newcommand{\pasj}{PASJ}

\title[NIR PLs of RRLs in the LMC]{The VMC Survey - XLII.  Near-infrared period-luminosity relations for RR Lyrae stars and the structure of the Large Magellanic Cloud.\thanks{Based on 
observations made with VISTA at ESO under programme ID 179.B-2003.}}
\author[F. Cusano et al.]{F. Cusano$^{1}$, M. I. Moretti $^{2}$, G. Clementini$^{1}$, V. Ripepi $^{2}$, M. Marconi$^{2}$,
\newauthor  M.-R. L. Cioni$^{3}$, S. Rubele$^{4}$, A. Garofalo$^{1}$, R. de Grijs$^{5,6,7}$,  M.~A.~T. Groenewegen$^8$,
\newauthor J.~M. Oliveira$^9$, S. Subramanian$^{10}$,  N.-C. Sun$^{11}$, J.~Th. van Loon$^9$
\\
$^{1}$ INAF-Osservatorio di Astrofisica e Scienza dello Spazio, Via Piero Gobetti, 93/3, I-40129 Bologna, Italy \\ 
$^{2}$ INAF-Osservatorio Astronomico di Capodimonte, Via Moiariello
 16, I-80131 Naples, Italy \\
$^{3}$ Leibniz-Institut f\"ur Astrophysik Potsdam, An der Sternwarte 16,D-14482 Potsdam, Germany\\
$^{4}$ INAF-Osservatorio Astronomico di Padova, vicolo dell'Osservatorio 5, I-35122  Padova, Italy \\
$^{5}$ Department of Physics and Astronomy, Macquarie University, Balaclava Road, Sydney, NSW 2109, Australia\\
$^{6}$ Research Centre for Astronomy, Astrophysics and Astrophotonics, Macquarie University, Balaclava Road, Sydney, NSW 2109, Australia\\
$^{7}$ International Space Science Institute--Beijing, 1 Nanertiao, Zhongguancun, Hai Dian District, Beijing 100090, China\\ 
$^8$ Koninklijke Sterrenwacht van Belgi\"e, Ringlaan 3, B--1180 Brussels,
Belgium \\
$^9$ Lennard-Jones Laboratories, School of Chemical and Physical Sciences, Keele University, ST5 5BG, UK \\
$^{10}$ Indian Institute of Astrophysics, 2nd Block, Bangalore - 560034, India \\
$^{11}$ The University of Sheffield, Western Bank, Sheffield, S10 2TN, UK
}

\begin{document}

\date{}

\pagerange{\pageref{firstpage}--\pageref{lastpage}} \pubyear{2020}

\maketitle 
\label{firstpage}

\begin{abstract}
We present results from an analysis of   $\sim$ 29,000 RR Lyrae stars 
located in the Large Magellanic Cloud (LMC). 
For these objects, near-infrared time-series photometry from the VISTA survey of the 
 Magellanic Clouds system (VMC) and optical   data from the OGLE  (Optical Gravitational Lensing Experiment) IV survey and the {\it Gaia} Data Release 2  catalogue of confirmed RR Lyrae stars were exploited. Using  VMC and OGLE~IV  magnitudes we  derived period--luminosity ({\it PL}), 
period--luminosity--metallicity ({\it PLZ}), period--Wesenheit ({\it PW}) and
period--Wesenheit--metallicity ({\it PWZ}) relations in all available bands. 
  More that ~7,000 RR Lyrae were discarded
from the analysis because they appear to be overluminous 
with respect to the PL relations.
The  $PL_{K_{\mathrm{s}}}$ 
relation was used to derive individual distance  to $\sim 22,000$ RR Lyrae stars,  and study the three-dimensional structure of the LMC.
The distribution of the LMC RR Lyrae stars is ellipsoidal with the three axis 
$S_1$=6.5 kpc, $S_2$=4.6 kpc  and $S_3$=3.7 kpc,  inclination  {\it i}=$22\pm4^{\circ }$  relative to the plane of the sky and   position angle of the line of nodes $\theta=167\pm7^{\circ }$
(measured from north to east).  The north-eastern part of the ellipsoid is closer to us 
and no particular associated  substructures  are detected as well as any metallicity gradient.

\end{abstract}


\section{Introduction}
Among the oldest objects in the Universe ($t>10$ Gyr)  RR Lyrae stars (RRLs) are pulsating variables with low masses (M$\leq$0.85 M$_{\odot}$)
and helium-burning cores,  which place them on the horizontal branch (HB) evolutionary sequence. 
RRLs are one of the most numerous type of pulsating stars, commonly found in globular clusters (GCs) and in galaxies hosting an old stellar component. 
The visual absolute magnitude of RRLs does not vary
significantly with period, because these variables are found on the HB, but has been found and predicted to depend on metallicity 
according to a luminosity--metallicity relation (M$_V$--[Fe/H]) which  is fundamental for the use of  RRLs as primary 
distance indicators (\citealt{vanAlbada&Baker1971}; \citealt{San81a,San81b}). 
On the other hand,  the dependence of the $V$--$K$ colour on  effective
temperature leads to the occurrence of 
a near-infrared (NIR)  
Period--Luminosity--Metallicity 
 ($PLZ$) relation for RRLs, first empirically observed by 
\cite{Lon86}.
As comprehensively discussed by \cite{Nem94},  the effect of reddening is
significantly reduced in the NIR compared  with optical wavelengths 
 \citep[A$_{Ks}$/A$_{V} \sim 1/9$, ][]{cardelli1989}  and the 
amplitude of pulsation is smaller than in the optical  
 \citep[Amp$_{Ks}$/Amp$_{V} \sim 1/4$, e. g. ][]{Braga2018},  allowing the derivation of  precise mean magnitudes even based on a
limited number of observations. 
This makes the RRL  $PLZ$  NIR  relations an even more powerful tool to measure distances than its optical counterparts. 
 Moreover, tighter relations are derived by introducing a colour term to the
$PL$ relation. In particular the Wesenheit function 
 \citep{vanden75, madore1982}
includes a color term whose coefficient is equal to the ratio of the total-to-selective extinction in a filter pair [$ W(X, Y)=  X - R_X \times (X - Y )$]. 
As such  the period-Wesenheit (PW) relation is reddening-free by definition.

The Large Magellanic Cloud (LMC)  is a dwarf irregular galaxy hosting several different stellar populations from 
old, to intermediate, to young, as well as a number of star forming regions. It is one of the closest companions of the Milky Way and the first rung of the astronomical distance ladder. Along with the Small Magellanic Cloud (SMC), the LMC is part of  the Magellanic System  which also comprises the Bridge connecting the two Clouds and the Magellanic Stream. The RRLs trace the oldest stellar component  and probe the structure of the LMC and SMC haloes, as opposed to  Classical Cepheids (CCs), which trace  regions of recent star formation like the LMC bar and spiral arms \citep{Mor14}. 

 The Magellanic RRLs have been observed in optical passbands and catalogued systematically by a number of different microlensing surveys:  the Massive Compact Halo Objects survey (MACHO; \citealt{Alc97}), the 
Optical Gravitational Lensing Experiment (OGLE;  \citealt{Uda97,Soszynski2016,Soszynski2019}), the Experi\'{e}nce pour la Recherche d'Objets Sombres 2 survey (EROS-2; \citealt{Tis07}) and, most recently, by the $Gaia$  mission (\citealt{prusti2016,brown2016,brown2018} and references therein). $Gaia$ is repeatedly monitoring the whole sky down to a limiting magnitude $G\sim$ 20.7--21 mag. 
By fully mapping  the Magellanic System
$Gaia$ is increasing the census of  RRLs,  
revealing the whole extension of the LMC and SMC haloes as well of
 the Bridge region (see, e.g. figs. 42 and 43 of  \citealt{Cle19}). About 2000 new RRLs published in the {\it Gaia} Data Release 2 (DR2; \citealt{Cle19}) were discovered in the Magellanic
 System; about 850 are outside and the remaining 1150 are inside  the OGLE footprint  as described by \citet{Soszynski2019}.
Nevertheless, the RRL catalogues published by 
OGLE \citep[e.g.][]{Sos09, Sos12, Soszynski2016, Soszynski2019} 
still represent fundamental references for the  Magellanic System  RRLs, since they provide light curves in the standard Johnson--Cousins $V_J, I_C$ passbands  
for about 48,000 of these variables, 
along with their pulsation period ({\it P}), mean magnitude and amplitude in the $I$ band,  and the main Fourier parameters of their light curves.

At  NIR wavelengths, an unprecedented step forward in our  knowledge of the  
properties of the  RRLs  
in the  Magellanic System  is being provided by 
the VISTA  $Y, J, K_\mathrm{s}$ 
survey of the Magellanic Clouds system (VMC\footnote{\url{http://star.herts.ac.uk/~mcioni/vmc}}, \citealt{Cio11}).  The survey aims at  studying the star formation history (SFH) and the three-dimensional (3-D) structure  of  the LMC, the SMC, the Bridge  and of a small section of the Magellanic Stream. 
The approach used to derive the SFH is based  on a combination of model single-burst populations that 
best describes the observed colour--magnitude diagrams (CMDs; see e.g., \citealt{Rub12,Rub15,Rub18}).  The 
 3-D geometry, instead,   is  inferred from  different  
distance indicators such as the luminosity of red clump stars 
\citep[RCs; see e.g.][]{subramanian2017} 
and the NIR  $PL$
relations of pulsating variable stars. 
Results for Magellanic System pulsating stars,  based on the VMC  
$K_\mathrm{s}$-band light curves,  
were presented by \cite{Rip12a} and \cite{Mur15, Mur18b} for RRLs; by   \cite{Rip12b,Rip16,Rip17} and \cite{Mor16} for CCs; by  \cite{Rip14,Rip15} 
 for anomalous and Type II Cepheids; 
whereas results for contact eclipsing binaries 
 observed by the VMC were presented by \cite{Mur14}.  A first comparative analysis of the 3-D structure of the LMC and SMC, as traced by RRLs, CCs and binaries separately, was presented by   \cite{Mor14}. In the optical a complete reconstruction of the 3-D LMC was performed by 
\citet{debandsingh2014} using  RRLs and  OGLE   data.

NIR PL relations for  RRLs in small regions of the LMC were obtained 
  by a few  authors before the present paper.
\citet{sze2008}  obtained deep NIR $J$ and $K$ band observations of six fields near the centre of the LMC. \citet{Bor09} investigated the metallicity dependence of the NIR PL in the $K$ band for 50 LMC RRLs. They found a very mild
dependence of the PL relation on metallicity.  \citet{Mur15}
presented  results from the analysis of 70 RRLs located in the bar of the LMC.
Combining spectroscopic metallicities  and VMC  $K_\mathrm{s}$ photometry, they
derived a new NIR PLZ relation.

In this paper we present results from an analysis of  
VMC $Y$-, $J$-, $K_\mathrm{s}$- and OGLE $V$-, $I$-band light 
curves of $\sim$ 29,000 LMC RRLs. 
 Optical and NIR data used in this study are presented in Section 2. 
 Section 3  describes in detail all  relevant steps  in our analysis to 
derive  the {\it PL(Z)} and {\it PW(Z)} relations  for the LMC RRLs. 
Section 4 presents the 3-D  structure of the LMC as traced by the 
RRLs. 
Finally, Section 5 summarises the results and  our main conclusions.

\section{data}\label{sec:data}
The observations of the VMC survey are 100 per cent complete with a total number of 110 fully completed 1.77 deg$^2$ VMC tiles, except for a few small gaps,
of which 68 cover the LMC, 13 are located in the Bridge, 27 cover the SMC, and two are 
in the Stream \citep[ see Figure 1 in][]{Cio17}.
The VMC survey was carried out  with the VISTA telescope which is located at Paranal Observatory 
in Chile. VISTA has a primary mirror of 4.1 m diameter at whose focus is placed  the 
VISTA InfraRed CAMera (VIRCAM) a wide-field infrared imager. For the VMC survey the 
$Y$, $J$ and $K_s$ filters were used.
To specifically enable the study of variable stars  the survey  obtained 
  $K_\mathrm{s}$ photometry in time-series mode  with  a minimum number of eleven deep epochs (each with 750 s exposure time per tile) and two 
 shallow epochs (each with 375 s exposure time per tile), which allow an optimal
sampling of the light curves for RRLs and 
for  CCs with periods of up to around 20--30 days, thus  providing the first 
comprehensive multi-epoch NIR  catalogue of  Magellanic System variables. Further details about the properties of the VMC data obtained for variable stars can be found in \cite{Mor14}.

The VMC images were processed by the Cambridge Astronomical Survey Unit (CASU) through the VISTA Data Flow System (VDFS) pipeline. 
The reduced data were then sent to the Wide Field
Astronomy Unit (WFAU) in Edinburgh where the single epochs were
stacked and catalogued by the Vista Science Archive (VSA; Lewis,
Irwin \& Bunclark 2010; Cross et al. 2012).
The VMC $K_\mathrm{s}$-band time-series reach a signal-to-noise ratio (S/N) of $~$5 at $K_\mathrm{s} \sim $19.3 mag. For comparison, the IRSF Near-Infrared Magellanic Clouds Survey \citep{Kat07} 
reaches $K_\mathrm{s} \sim$ 17.0 mag for the same S/N level.  
The sensitivity and limiting magnitude achieved by the VMC survey allow us to obtain  $K_{\mathrm{s}}$ light curves 
for RRLs in both Magellanic Clouds (MCs).  
Furthermore, the 
large area coverage of the VMC enables, for the first time, to systematically study the NIR $PL$ relation 
 of RRLs across the whole  Magellanic System.

\begin{figure}
\begin{centering}
\includegraphics[width=8.6cm, height=8.6cm, clip]{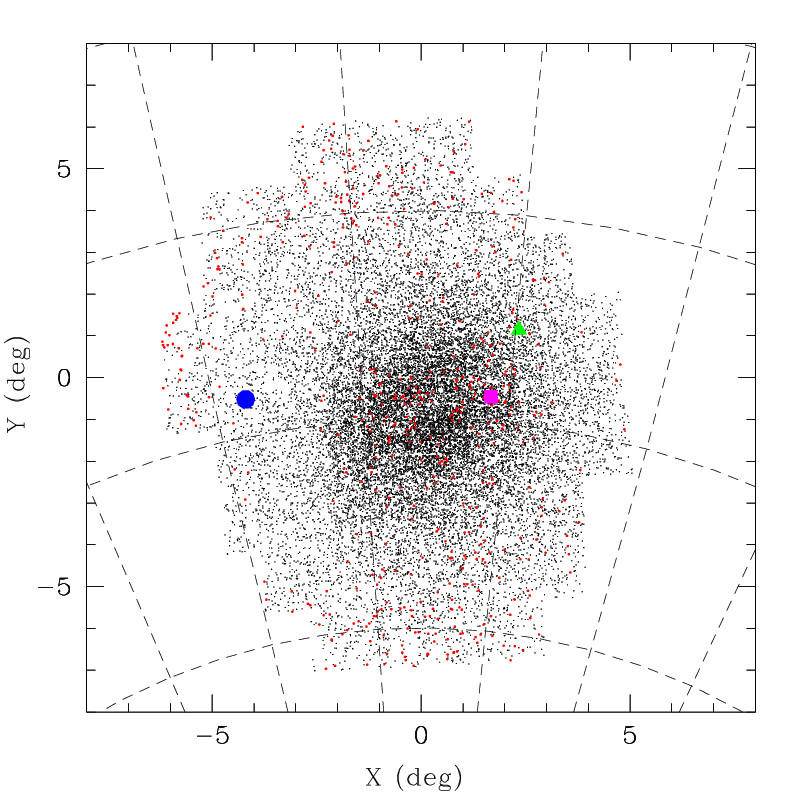}
\caption{VMC coverage of the LMC. Black dots mark the $\sim 22,000$ RRLs analysed in
this paper.  Red dots are RRLs present in the $Gaia$ catalogue only. 
The centre of the map is $\alpha_0$=81.0 deg and $\delta_0$=$-$69.0 deg.
The blue circle,   magenta square and green triangle indicate the GCs  
NGC~2210, NGC~1835 and NGC~1786, respectively  (see Section 4.3).}\label{fig:RR_LMCmap}
\end{centering}
\end{figure} 
We cross-matched the VMC catalogue with the OGLE catalogue of RRLs in the LMC \citep{Soszynski2016,Soszynski2019}  and with the catalogue of confirmed LMC RRLs published as part of $Gaia$ DR2 \citep{Cle19}.  We used $Gaia$ DR2 as the more recent  eDR3
\citep{brown2020}
does not contain an updated list of variable sources. This will become available with DR3 in 2022. The total number of OGLE RRLs in the VMC footprint is $\sim 35,000$, but  in our analysis
we considered  only the OGLE  fundamental-mode (RRab) and first-overtone pulsators (RRc): (i)
for which  both $V$ and $I$ light curves are available,
(ii) without flags in the OGLE catalogue such as: 
uncertain, secondary period, possible Anomalous Cepheid, possible $\delta$ Scuti star, etc.,
and (iii) with a VMC counterpart found within 1 arcsec.   
Our total sample thus selected included   28,132 RRLs (21,152 RRab and 6,980 RRc).
Furthermore,  we cross-matched the {\it Gaia} catalogue of confirmed RRLs
with the general catalogue of VMC sources in the LMC field of view.
This allowed us to recover an additional 524 RRLs (357 RRab, and  167 RRc) which were missing in the OGLE catalogue of LMC RRLs. 
$Gaia$ has a smoother spatial sampling than OGLE thus allowing to recover additional
RRLs in regions near to  CCD edges and/or across inter-CCD gaps of the OGLE camera. 

Figure~\ref{fig:RR_LMCmap} shows the  spatial distribution of the 
full sample of RRLs analysed in this work.
The {\it X} and {\it Y} coordinates are defined as in \cite{VC01} with
$\alpha_0$=81.0 deg and $\delta_0$=$-$69.0 deg. 
 Table~\ref{tab:matching} provides centre coordinates for the 68 VMC tiles covering the LMC  and the number of RRLs contained  in each tile.

 \linespread{1}
\begin{table*}
\caption{VMC tiles in the LMC analysed in the present study: (1) field and tile number; (2), (3) coordinates of the tile centre, J2000 epoch; 
(4) total number of RRLs in the tile, 
(5) extinction $E(V-I)$ and (6) absorption in the $K_{\mathrm{s}}$ band  (see Section~3.3);
(7) average period of the RRab stars in the tile and  (8) dispersion of the average period of the RRab stars (see Section 2); (9), (10), (11), (12) and (13)
coefficients of the $PL$ relations (in the form: $K_{\mathrm{s}_0}$=$a$ + $b$ $\times\log P$) for RRab and fundamentalised (by adding 0.127 to the logarithm of the first overtone period) RRc stars and their uncertainties (see Section 3.4).}
\label{tab:matching}
\tiny
\begin{tabular}{l|c|c|c|c|c|c|l|c|c|c|l|l}
\hline
 Tile      &     R.A.          &   Dec                     & N~~~       & $\langle E(V-I)\rangle$  &     $\langle AK_{\mathrm{s}}\rangle$  &  $\langle Pab \rangle$ & $\sigma_{\rm Pab}$  &  $a$   & $b$  &  $\sigma_{ a}$     &    $\sigma_{ b}$    &     rms  \\  
        &($^h$:$^m$:$^s$)  &($^{\circ}$:$^{\prime}$:$^{\prime\prime}$) &  &  (mag)    &  (mag)  & (d)     &   (d)  &   (mag)       &  (mag)      &  (mag)   & (mag)    &    \\
 \hline
LMC 2\_3     &    04:48:04 &   $-$74:54:11	  &    84      &    0.121    & 0.035	&   0.577   &	   0.067 & 17.45  &   $-$2.56    &   0.05	 &  0.17  &  0.11    \\ 
LMC 2\_4     &    05:04:43 &   $-$75:04:45	  &   111      &    0.134    & 0.039	&   0.586   &	   0.061 & 17.50  &   $-$2.34    &   0.04	 &  0.15  &  0.11    \\ 
LMC 2\_5     &    05:21:39 &   $-$75:10:50	  &   125      &    0.158    & 0.046	&   0.575   &	   0.066 & 17.44  &   $-$2.41    &   0.04	 &  0.17  &  0.12    \\ 
LMC 2\_6     &    05:38:43 &   $-$75:12:21	  &   107      &    0.128    & 0.037	&   0.596   &	   0.082 & 17.38  &   $-$2.68    &   0.04	 &  0.17  &  0.13    \\ 
LMC 2\_7     &    05:55:46 &   $-$75:09:17	  &    82      &    0.145    & 0.042	&   0.577   &	   0.066 & 17.43  &   $-$2.35    &   0.06	 &  0.21  &  0.12    \\ 
LMC 3\_2     &    04:37:05 &   $-$73:14:30	  &   114      &    0.105    & 0.031	&   0.594   &	   0.064 & 17.46  &   $-$2.53    &   0.04	 &  0.14  &  0.10    \\ 
LMC 3\_3     &    04:52:00 &   $-$73:28:09	  &   170      &    0.111    & 0.032	&   0.587   &	   0.069 & 17.43  &   $-$2.65    &   0.03	 &  0.11  &  0.11    \\ 
LMC 3\_4     &    05:07:14 &   $-$73:37:50	  &   216      &    0.127    & 0.037	&   0.580   &	   0.063 & 17.45  &   $-$2.55    &   0.03	 &  0.11  &  0.11    \\ 
LMC 3\_5     &    05:22:43 &   $-$73:43:25	  &   250      &    0.121    & 0.035	&   0.589   &	   0.072 & 17.42  &   $-$2.59    &   0.03	 &  0.10  &  0.12    \\ 
LMC 3\_6     &    05:38:18 &   $-$73:44:51	  &   218      &    0.133    & 0.038	&   0.578   &	   0.064 & 17.42  &   $-$2.54    &   0.03	 &  0.12  &  0.12    \\ 
LMC 3\_7     &    05:53:52 &   $-$73:42:06	  &   160      &    0.115    & 0.033	&   0.579   &	   0.082 & 17.45  &   $-$2.39    &   0.03	 &  0.12  &  0.12    \\ 
LMC 3\_8     &    06:09:17 &   $-$73:35:12	  &   138      &    0.118    & 0.034	&   0.573   &	   0.059 & 17.38  &   $-$2.51    &   0.04	 &  0.16  &  0.12    \\ 
LMC 4\_2     &    04:41:31 &   $-$71:49:16	  &   232      &    0.139    & 0.040	&   0.581   &	   0.062 & 17.46  &   $-$2.52    &   0.03	 &  0.11  &  0.11    \\ 
LMC 4\_3     &    04:55:19 &   $-$72:01:53	  &   311      &    0.110    & 0.032	&   0.584   &	   0.068 & 17.45  &   $-$2.62    &   0.02	 &  0.09  &  0.10    \\ 
LMC 4\_4     &    05:09:24 &   $-$72:10:50	  &   485      &    0.080    & 0.023	&   0.578   &	   0.067 & 17.47  &   $-$2.46    &   0.02	 &  0.08  &  0.12    \\ 
LMC 4\_5     &    05:23:40 &   $-$-72:16:00       &   522      &    0.082    & 0.024	&   0.582   &	   0.077 & 17.42  &   $-$2.60    &   0.02	 &  0.07  &  0.12    \\ 
LMC 4\_6     &    05:38:00 &   $-$72:17:20	  &   431      &    0.097    & 0.028	&   0.576   &	   0.066 & 17.44  &   $-$2.46    &   0.02	 &  0.09  &  0.13    \\ 
LMC 4\_7     &    05:52:20 &   $-$72:14:50	  &   262      &    0.105    & 0.031	&   0.578   &	   0.066 & 17.43  &   $-$2.43    &   0.03	 &  0.11  &  0.12    \\ 
LMC 4\_8     &    06:06:33 &   $-$72:08:31	  &   143      &    0.102    & 0.030	&   0.581   &	   0.070 & 17.37  &   $-$2.67    &   0.04	 &  0.14  &  0.11    \\ 
LMC 4\_9     &    06:20:33 &   $-$71:58:27	  &   122      &    0.078    & 0.023	&   0.589   &	   0.076 & 17.27  &   $-$2.90    &   0.04	 &  0.15  &  0.12    \\ 
LMC 5\_1     &    04:32:44 &   $-$70:08:40	  &   154      &    0.109    & 0.032	&   0.593   &	   0.068 & 17.42  &   $-$2.71    &   0.03	 &  0.10  &  0.10    \\ 
LMC 5\_2     &    04:45:19 &   $-$70:23:44	  &   288      &    0.149    & 0.043	&   0.584   &	   0.071 & 17.50  &   $-$2.35    &   0.03	 &  0.10  &  0.12    \\ 
LMC 5\_3     &    04:58:12 &   $-$70:35:28	  &   618      &    0.086    & 0.025	&   0.583   &	   0.067 & 17.46  &   $-$2.57    &   0.02	 &  0.07  &  0.13    \\ 
LMC 5\_4     &    05:11:17 &   $-$70:43:46	  &  1138      &    0.082    & 0.02	&   0.582   &	   0.069 & 17.44  &   $-$2.60    &   0.02	 &  0.06  &  0.15    \\ 
LMC 5\_5     &    05:24:30 &   $-$70:48:34	  &  1349      &    0.085    & 0.024	&   0.578   &	   0.070 & 17.47  &   $-$2.42    &   0.02	 &  0.06  &  0.16    \\ 
LMC 5\_6     &    05:37:48 &   $-$70:49:49	  &   946      &    0.136    & 0.039	&   0.575   &	   0.072 & 17.43  &   $-$2.52    &   0.02	 &  0.08  &  0.17    \\ 
LMC 5\_7     &    05:51:05 &   $-$70:47:31	  &   512      &    0.133    & 0.039	&   0.574   &	   0.067 & 17.41  &   $-$2.53    &   0.02	 &  0.09  &  0.14    \\ 
LMC 5\_8     &    06:04:16 &   $-$70:41:40	  &   190      &    0.084    & 0.024	&   0.580   &	   0.065 & 17.41  &   $-$2.48    &   0.03	 &  0.10  &  0.11    \\ 
LMC 5\_9     &    06:17:18 &   $-$70:32:21	  &   151      &    0.088    & 0.025	&   0.585   &	   0.066 & 17.42  &   $-$2.43    &   0.03	 &  0.12  &  0.10    \\ 
LMC 6\_1     &    04:36:49 &   $-$68:43:51	  &   213      &    0.084    & 0.024	&   0.576   &	   0.069 & 17.43  &   $-$2.59    &   0.03	 &  0.11  &  0.13    \\ 
LMC 6\_2     &    04:48:39 &   $-$68:57:56	  &   352      &    0.137    & 0.040	&   0.577   &	   0.070 & 17.45  &   $-$2.54    &   0.03	 &  0.09  &  0.13    \\ 
LMC 6\_3     &    05:00:42 &   $-$69:08:54	  &   776      &    0.090    & 0.026	&   0.578   &	   0.072 & 17.47  &   $-$2.53    &   0.02	 &  0.08  &  0.17    \\ 
LMC 6\_4     &    05:12:56 &   $-$69:16:39	  &  1305      &    0.094    & 0.027	&   0.581   &	   0.074 & 17.47  &   $-$2.47    &   0.02	 &  0.07  &  0.19    \\ 
LMC 6\_5     &    05:25:16 &   $-$69:21:08	  &  1294      &    0.100    & 0.029	&   0.580   &	   0.071 & 17.52  &   $-$2.35    &   0.02	 &  0.07  &  0.20    \\ 
LMC 6\_6     &    05:37:40 &   $-$69:22:18	  &   978      &    0.197    & 0.057	&   0.575   &	   0.071 & 17.41  &   $-$2.68    &   0.02	 &  0.08  &  0.18    \\ 
LMC 6\_7     &    05:50:03 &   $-$69:20:09  &   436      &    0.194    & 0.056	&   0.581   &	   0.075 & 17.36  &   $-$2.64    &   0.02	 &  0.09  &  0.14    \\ 
LMC 6\_8     &    06:02:21 &   $-$69:14:42	  &   190      &    0.059    & 0.017	&   0.585   &	   0.064 & 17.37  &   $-$2.55    &   0.03	 &  0.11  &  0.11    \\ 
LMC 6\_9     &    06:14:33 &   $-$69:06:00	  &   129      &    0.057    & 0.017	&   0.580   &	   0.067 & 17.33  &   $-$2.54    &   0.03	 &  0.12  &  0.10    \\ 
LMC 6\_10    &    06:26:32 &   $-$68:54:06	  &   103      &    0.073    & 0.021	&   0.585   &	   0.066 & 17.30  &   $-$2.68    &   0.04	 &  0.14  &  0.10    \\ 
LMC 7\_1     &    04:40:09 &   $-$67:18:20        &   125      &    0.070    & 0.020	&   0.585   &	   0.077 & 17.45  &   $-$2.52    &   0.04	 &  0.15  &  0.12    \\ 
LMC 7\_2     &    04:51:35 &   $-$67:31:57	  &   291      &    0.097    & 0.028	&   0.584   &	   0.072 & 17.44  &   $-$2.57    &   0.03	 &  0.10  &  0.13    \\ 
LMC 7\_3     &    05:02:55 &   $-$67:42:15	  &   477      &    0.089    & 0.026	&   0.572   &	   0.072 & 17.43  &   $-$2.62    &   0.02	 &  0.08  &  0.14    \\ 
LMC 7\_4     &    05:14:24 &   $-$67:49:31	  &   749      &    0.102    & 0.030	&   0.577   &	   0.072 & 17.47  &   $-$2.43    &   0.02	 &  0.07  &  0.14    \\ 
LMC 7\_5     &    05:25:58 &   $-$67:53:42	  &   709      &    0.137    & 0.040	&   0.579   &	   0.073 & 17.44  &   $-$2.50    &   0.02	 &  0.07  &  0.13    \\ 
LMC 7\_6     &    05:37:35 &   $-$67:54:47	  &   440      &    0.116    & 0.034	&   0.582   &	   0.069 & 17.42  &   $-$2.46    &   0.02	 &  0.08  &  0.13    \\ 
LMC 7\_7     &    05:49:12 &   $-$67:52:45	  &   265      &    0.113    & 0.033	&   0.581   &	   0.064 & 17.35  &   $-$2.57    &   0.03	 &  0.10  &  0.12    \\ 
LMC 7\_8     &    06:00:45 &   $-$67:47:38	  &   162      &    0.056    & 0.016	&   0.598   &	   0.074 & 17.38  &   $-$2.38    &   0.03	 &  0.11  &  0.11    \\ 
LMC 7\_9     &    06:12:12 &   $-$67:39:26	  &   100      &    0.051    & 0.015	&   0.579   &	   0.075 & 17.32  &   $-$2.65    &   0.03	 &  0.12  &  0.10    \\ 
LMC 7\_10    &    06:23:29 &   $-$67:28:15	  &    48      &    0.061    & 0.018	&   0.599   &	   0.078 & 17.40  &   $-$2.23    &   0.06	 &  0.22  &  0.11    \\ 
LMC 8\_2     &    04:54:12 &   $-$66:05:48	  &   186      &    0.104    & 0.030	&   0.593   &	   0.071 & 17.43  &   $-$2.58    &   0.03	 &  0.12  &  0.12    \\ 
LMC 8\_3     &    05:04:54 &   $-$66:15:30	  &   238      &    0.071    & 0.020	&   0.580   &	   0.071 & 17.38  &   $-$2.62    &   0.03	 &  0.12  &  0.13    \\ 
LMC 8\_4     &    05:15:43 &   $-$66:22:20	  &   301      &    0.095    & 0.028	&   0.582   &	   0.071 & 17.42  &   $-$2.45    &   0.03	 &  0.10  &  0.13    \\ 
LMC 8\_5     &    05:26:38 &   $-$66:26:16	  &   287      &    0.086    & 0.025	&   0.585   &	   0.075 & 17.38  &   $-$2.56    &   0.03	 &  0.10  &  0.12    \\ 
LMC 8\_6     &    05:37:34 &   $-$66:27:16	  &   255      &    0.082    & 0.024	&   0.586   &	   0.067 & 17.37  &   $-$2.50    &   0.03	 &  0.11  &  0.13    \\ 
LMC 8\_7     &    05:48:30 &   $-$66:25:20	  &   172      &    0.073    & 0.021	&   0.575   &	   0.066 & 17.34  &   $-$2.54    &   0.04	 &  0.14  &  0.13    \\ 
LMC 8\_8     &    05:59:23 &   $-$66:20:29	  &   145      &    0.040    & 0.012	&   0.596   &	   0.077 & 17.36  &   $-$2.52    &   0.03	 &  0.11  &  0.10    \\ 
LMC 8\_9     &    06:10:11 &   $-$66:12:44	  &   101      &    0.062    & 0.018	&   0.582   &	   0.071 & 17.30  &   $-$2.54    &   0.04	 &  0.16  &  0.11    \\ 
LMC 9\_3     &    05:06:41 &   $-$64:48:40	  &   111      &    0.052    & 0.015	&   0.594   &	   0.067 & 17.45  &   $-$2.30    &   0.04	 &  0.15  &  0.12    \\ 
LMC 9\_4     &    05:16:55 &   $-$64:55:08	  &   178      &    0.061    & 0.018	&   0.581   &	   0.067 & 17.38  &   $-$2.46    &   0.03	 &  0.11  &  0.11    \\ 
LMC 9\_5     &    05:27:14 &   $-$64:58:49	  &   209      &    0.072    & 0.021	&   0.582   &	   0.063 & 17.38  &   $-$2.40    &   0.03	 &  0.10  &  0.11    \\ 
LMC 9\_6     &    05:37:35 &   $-$64:59:44  &   163      &    0.059    & 0.017	&   0.582   &	   0.063 & 17.28  &   $-$2.67    &   0.03	 &  0.12  &  0.11    \\ 
LMC 9\_7     &    05:47:55 &   $-$64:57:53	  &   117      &    0.068    & 0.020	&   0.589   &	   0.065 & 17.35  &   $-$2.37    &   0.04	 &  0.14  &  0.11    \\ 
LMC 9\_8     &    05:58:13 &   $-$64:53:15	  &   114      &    0.047    & 0.014	&   0.588   &	   0.071 & 17.31  &   $-$2.49    &   0.04	 &  0.14  &  0.10    \\ 
LMC 9\_9     &    06:08:26 &   $-$64:45:53	  &    61      &    0.083    & 0.024	&   0.568   &	   0.062 & 17.23  &   $-$2.81    &   0.06	 &  0.20  &  0.12    \\ 
LMC 10\_4    &    05:18:01 &   $-$63:27:54	  &   107      &    0.045    & 0.013	&   0.583   &	   0.070 & 17.37  &   $-$2.49    &   0.03	 &  0.12  &  0.09    \\ 
LMC 10\_5    &    05:27:49 &   $-$63:31:23	  &   116      &    0.061    & 0.018	&   0.587   &	   0.065 & 17.33  &   $-$2.59    &   0.04	 &  0.14  &  0.11    \\ 
LMC 10\_6    &    05:37:38 &   $-$63:32:13	  &   105      &    0.077    & 0.022	&   0.585   &	   0.066 & 17.36  &   $-$2.38    &   0.04	 &  0.14  &  0.11    \\ 
LMC 10\_7    &    05:47:26 &   $-$63:30:25	  &    91      &    0.050    & 0.015	&   0.581   &	   0.070 & 17.32  &   $-$2.53    &   0.05	 &  0.19  &  0.12    \\ 

\hline
\end{tabular}
\medskip
\end{table*}
\linespread{1.3}

Time-series  $Y$, $J$ and $K_\mathrm{s}$ photometry for the RRLs observed by the VMC survey is provided 
in Table~\ref{tab:lc}. 
The table contains  the Heliocentric Julian Date (HJD) of the observations inferred from the VSA (column 1), the $Y$, $J$ and 
$K_\mathrm{s}$  magnitudes 
obtained 
using an aperture photometry diameter of 2.0 arcsec (column 2), and the error on the magnitudes (column 3). The light curves contain only epoch data observed within pre-defined constraints, 
 that is with conditions of seeing $< 1^{\prime\prime}$ for $K_\mathrm{s}$, $< 1.2^{\prime\prime}$ for $Y$ and, $< 1.1^{\prime\prime}$ for $J$, airmass $<$1.7 
and moon distance $> 70^o$ \citep{Cio11}. 
 The full catalogue of light curves 
is available in the electronic version of the paper. 

\linespread{1}
\begin{table}
\caption{$Y$, $J$ and $K_\mathrm{s}$ time-series  photometry for the RRLs analysed in this paper. The table is published in its entirety as Supporting Information with the electronic version of the article. 
A portion is shown here for guidance regarding its format and content.
}
\label{tab:lc}
\begin{center}
\begin{tabular}{c|c|c}
\hline
\hline
\multicolumn{3}{c}{{ VMC-558396536162}}\\
\hline
    HJD$-$2400000      &    $Y$    & err$_Y$ \\
\hline
   55583.680443  &   18.24    &  0.02\\
   55597.606813  &   18.42    &  0.03\\
   55878.818319  &   18.39    &  0.03\\
   55983.600130  &   18.30    &  0.02\\
   56004.513472  &   18.51    &  0.03\\
\hline
    HJD$-$2400000      &    $J$    & err$_J$ \\
\hline
   55589.666983   &  18.07    &  0.02\\
   55615.569205   &  18.13    &  0.03\\
   55980.563227   &  18.10    &  0.03\\
   56004.533138   &  18.22    &  0.04\\
\hline
    HJD$-$2400000      &    $K_\mathrm{s}$    & err$_{K\mathrm{s}}$ \\
\hline
   55979.576519  &   17.73    &  0.05\\
   55980.586298  &   17.78    &  0.08\\
   55981.562849  &   17.71    &  0.05\\
   55983.620398  &   17.78    &  0.07\\
   55986.562678  &   17.83    &  0.06\\
   55992.575373  &   17.80    &  0.07\\
   56009.519064  &   17.76    &  0.07\\
   56178.854539  &   17.78    &  0.05\\
   56197.794262  &   17.86    &  0.06\\
   56214.835096  &   17.89    &  0.06\\
   56232.750233  &   17.77    &  0.05\\
   56232.792530  &   17.76    &  0.05\\
   56255.668172  &   17.89    &  0.06\\
   57044.537788  &   18.05    &  0.13\\
\hline
\end{tabular}
\end{center}
\medskip
\end{table}
\linespread{1.3}

\begin{figure}
\begin{centering}
\includegraphics[width=9cm, height=6cm]{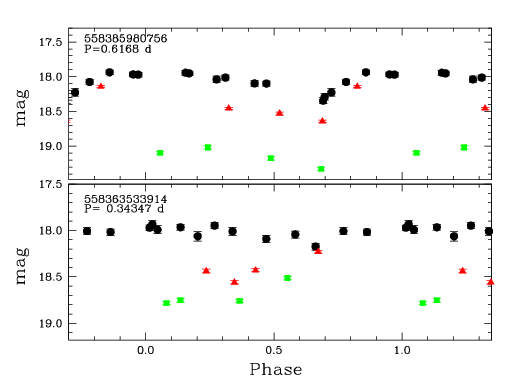}
\caption{$Y$ (green squares), $J$ (red triangles) and $K_\mathrm{s}$ (black dots)  light curves for an RRab (top panel) and an RRc (bottom panel) star in the LMC field.  The VMC
ID and the period are provided in the top left-hand corner.
Typical errors for individual  $K_\mathrm{s}$ epochs are  about 0.06 mag  and 0.1 mag for RRab and RRc stars, respectively. For $Y$- and $J$-bands typical errors
are 0.03 and 0.07 mag for RRab and RRc stars, respectively.}\label{fig:lc_abk}
\end{centering}
\end{figure}

The $K_{\mathrm{s}}$  photometry of the VISTA system is tied to the Two Micron All Sky Survey (2MASS;  \citealt{Skr06}) photometry.  
The complete set of transformation relations from one system to the other is available on the CASU web 
site\footnote{\url{http://casu.ast.cam.ac.uk/surveys-projects/vista/technical/photometric-properties}}  or in the work by \citet{gonz2018}. 
We used   the photometry in the VISTA system v1.5. The periods provided by the OGLE  catalogue were used to fold the $Y, J$ and $K_\mathrm{s}$ light curves of the RRLs observed by OGLE, whereas we used the periods provided in the  {\it Gaia} DR2 RRL {\it vari}\_table for the sources observed only by {\it Gaia} as  the eDR3 release
does not contain update information for variable stars like period, 
epoch of  maximum light and photometry.
Examples of the  $Y$, $J$  and $K_\mathrm{s}$ light curves for an RRab and an RRc star are shown in Figure~\ref{fig:lc_abk}. A complete atlas of light curves is  provided in electronic form.
 The  $K_\mathrm{s}$
light curves are very well sampled, with on average, 13--14 sampling points,
confirming the soundness of the VMC observing strategy for RRLs. 
 Typical errors for the 
individual $K_\mathrm{s}$ data points are in the range of 0.06 -- 0.1 mag.
 The light curves  in the $Y$ and $J$ bands have on average four  phase points. 
Average errors for the individual $Y$ and $J$ observations are 0.03 and 0.04 mag, respectively.

\linespread{1}
\begin{table*}
\tiny
\caption{List of LMC RRLs analysed in the paper:
OGLE~IV ID (5 digit number 
$-$OGLE-LMC-RRLYR-) or {\it Gaia} DR2 ID ($>$9 digit number) column (1),
 VMC ID is column (2),
 J2000 coordinates R.A. and Dec. from the VMC catalog are columns (3) and (4),
pulsation mode (column 5), period (column 6) and  epoch of maximum light ($-$2456000 for OGLE sources, $-$  2455197.5  for {\it Gaia} DR2 sources) column (7), metallicity is column (8),
 absorption in the  $K_\mathrm{s}$-band (column 9), number of epochs in the $K_\mathrm{s}$-band (column 10),  Y (column 11) and J band (column 12),
columns (13), (14), and (15) are average VMC magnitudes: $\langle Y \rangle$, $\langle J \rangle$ and $\langle K_{\mathrm{s}} \rangle$. 
The table is published in its entirety as Supporting Information with the electronic version of the article.  
A portion is shown here for guidance regarding its form and content.}\label{tab:RR88}
\begin{center}
\begin{tabular}{|l|l|c|c|c|c|c|c|c|c|c|c|c|c|c|}
\hline
\hline
 ID &   ID(VMC) & R.A.  & Dec. & Type    & $P$             &  $T_0$  & [Fe/H]  &  $A_{K_\mathrm{s}}$      &   $N_{K_{\mathrm{s}}}$   	& $N_{Y}$   	& $N_{J}$   	&   $\langle K_\mathrm{s} \rangle$  & $\langle Y \rangle$  &$\langle J \rangle$    \\		  
&  & deg & deg     &              &            (d)     & (d) & dex       &     (mag)      &           &        &  &   (mag)       &    (mag)
       &  (mag)    \\
\hline
 34314 & 558346514686 & 84.202208  & --74.490389 &  ab  &      0.7040914  &  0.46883   &           &0.028  & 26 &    8 & 8 &18.12 &18.64  &  18.50     \\
 33596 & 558346517950 & 82.979042  & --74.501333 &  ab  &      0.6610381  &  0.29199   & $-$1.424  &0.035  & 13 &    4 & 4 &17.94 &18.39  &  18.21     \\
 35139 & 558346518034 & 85.807625  & --74.498722 &  ab  &      0.4889976  &  0.25714   & $-$1.592  &0.036  & 13 &    4 & 4 &18.20 &18.88  &  18.66     \\
 34860 & 558346519807 & 85.190208  & --74.510083 &  c   &      0.3114614  &  0.24308   & $-$1.507  &0.082  & 13 &    4 & 4 &18.45 &18.88  &  18.77     \\
 34267 & 558346526561 & 84.127458  & --74.540444 &  ab  &      0.53476    &  0.08816   &           &0.037  & 13 &    4 & 4 &18.10 &18.53  &  18.32     \\
 34756 & 558346526744 & 85.008125  & --74.538944 &  ab  &      0.55045    &  0.16538   &           &0.082  & 12 &    3 & 4 &17.95 &18.37  &  18.25     \\
 34635 & 558346527385 & 84.747875  & --74.543194 &  c   &      0.34962    &  0.01419   & $-$1.527  &0.043  & 13 &    4 & 4 &18.41 &18.82  &  18.60     \\
\hline				 	 						  
\end{tabular}
\end{center}
\end{table*}
\linespread{1.3}

\section{Period--Luminosity and Period--Wesenheit relations}\label{sec:PLk}
\subsection{Average NIR magnitudes }
Mean $Y, J$ and $K_\mathrm{s}$ magnitudes for the LMC RRLs 
were derived as a simple mean of magnitudes expressed in flux units, without modelling the light curves. 

For RRLs observed in the optical,  integrating the modelled light curve over the whole pulsation cycle is the correct procedure to derive the mean magnitude. However, in the NIR, amplitudes are so small that the difference between intensity-averaged mean magnitudes of the modelled light curves and the simple mean of the magnitudes in flux units is negligible.  
A comparison of the two mean values, where the former were 
computed  by modelling the light curves with a Fourier series 
using the Graphical Analyzer
of Time Series package (GRaTIS; custom
software developed at the Bologna Observatory by
P. Montegriffo, see e.g. \citealt{Cle00}), for a sample of 170 LMC RRLs showed  that the differences between the two procedures are less than  0.02 mag, that is, smaller than  the average errors  of the two methods. 
Table~\ref{tab:RR88} 
lists the  $Y$, $J$ and $K_\mathrm{s}$ mean magnitudes 
obtained as simple means of the magnitudes (in flux units)  along with the main properties of
the stars, namely,  OGLE~IV or {\it Gaia} DR2 ID, period, pulsation mode, epoch of maximum light, VMC ID
and  number of observations in the $K_\mathrm{s}$, $Y$ and $J$ bands.

\begin{figure}
\begin{centering}
\includegraphics[width=8cm, height=8cm]{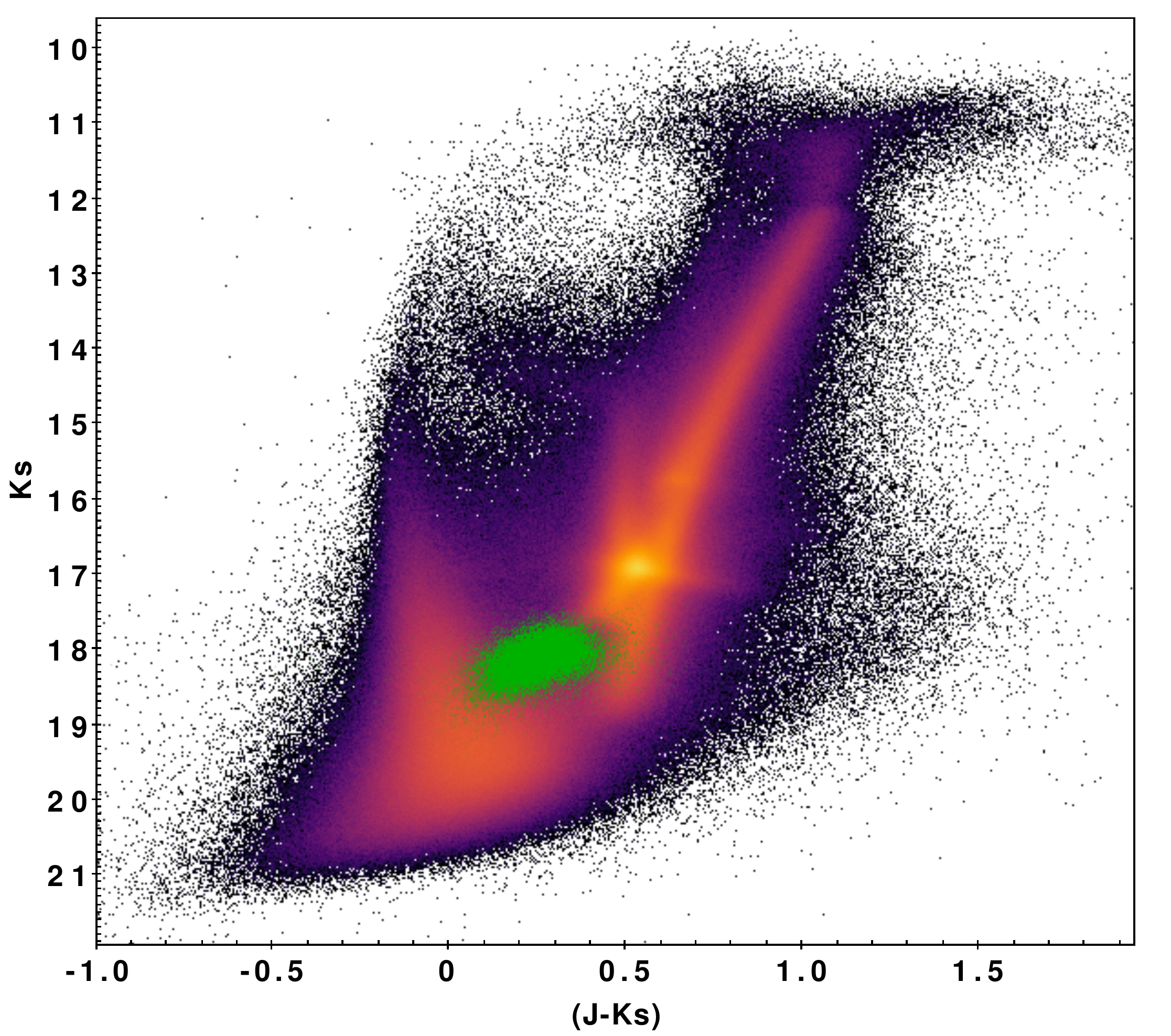}
\caption{CMD of all  VMC sources matching  the  {\it Gaia} eDR3 catalogue using the criteria defined in the text. A total  of ~6.9 million sources are shown 
in this diagram. The RRLs analysed in this paper are marked by small green dots.}\label{fig:cmd}
\end{centering}
\end{figure}

The average magnitudes  were used to place the RRLs in the $ J - K_\mathrm{s}$ vs $K_\mathrm{s}$ CMD,  shown  in Figure~\ref{fig:cmd}. 
The RRLs are marked with small green dots. 
 To construct the CMD we selected from  the {\it Gaia} eDR3 catalogue 
\citep{brown2020} 
  an area of 8.8 degrees in radius around the centre of the LMC ($\alpha_0$=80.8 deg, $\delta_0$=$-$69.7 deg) which contained more than  25 million objects.  We chose this  area in order to fully cover the LMC regions tiled by the VMC observations  as well as the whole footprint of the OGLE-IV survey. 
 We assumed as bona fide LMC members only sources in the selected area with parallaxes ($\varpi$) and proper motions ($\mu_{\alpha}^{*}, \mu_{\delta}$) satisfying the following criterion:
\begin{equation}
\begin{split}
\sqrt((\varpi-\varpi_{\rm LMC})^{2}+ (\mu_{\alpha}^{*}- {\mu_{\alpha}^{*}}_{\rm LMC})^{2}+\\
    (\mu_{\delta}- {\mu_{\delta}}_{\rm LMC})^{2})<1, 
\end{split}
\end{equation}
 where $\varpi_{\rm LMC}=-0.0040 \pm 0.3346$ mas , ${\mu_{\alpha}^{*}}_{\rm LMC}=1.7608 \pm 0.4472$ mas yr$^{-1}$ and ${\mu_{\delta}}_{\rm LMC}= 0.3038 \pm 0.6375$  mas yr$^{-1}$ are the mean LMC parallax and proper motions  from  \citet{luri2020}.
We then cross-matched the sources satisfying the above condition with the VMC general catalogue, thus obtaining a final sample of $\sim 6.9\times10^6$ sources shown in 
Figure~\ref{fig:cmd}. The proper motions and parallaxes of the CMD stars are  from $Gaia$ eDR3  which contains the most recent and accurate astrometry and photometry available from $Gaia$. 

\linespread{1}
\begin{figure}
\begin{center}
\includegraphics[width=8.6cm, height=8.6cm]{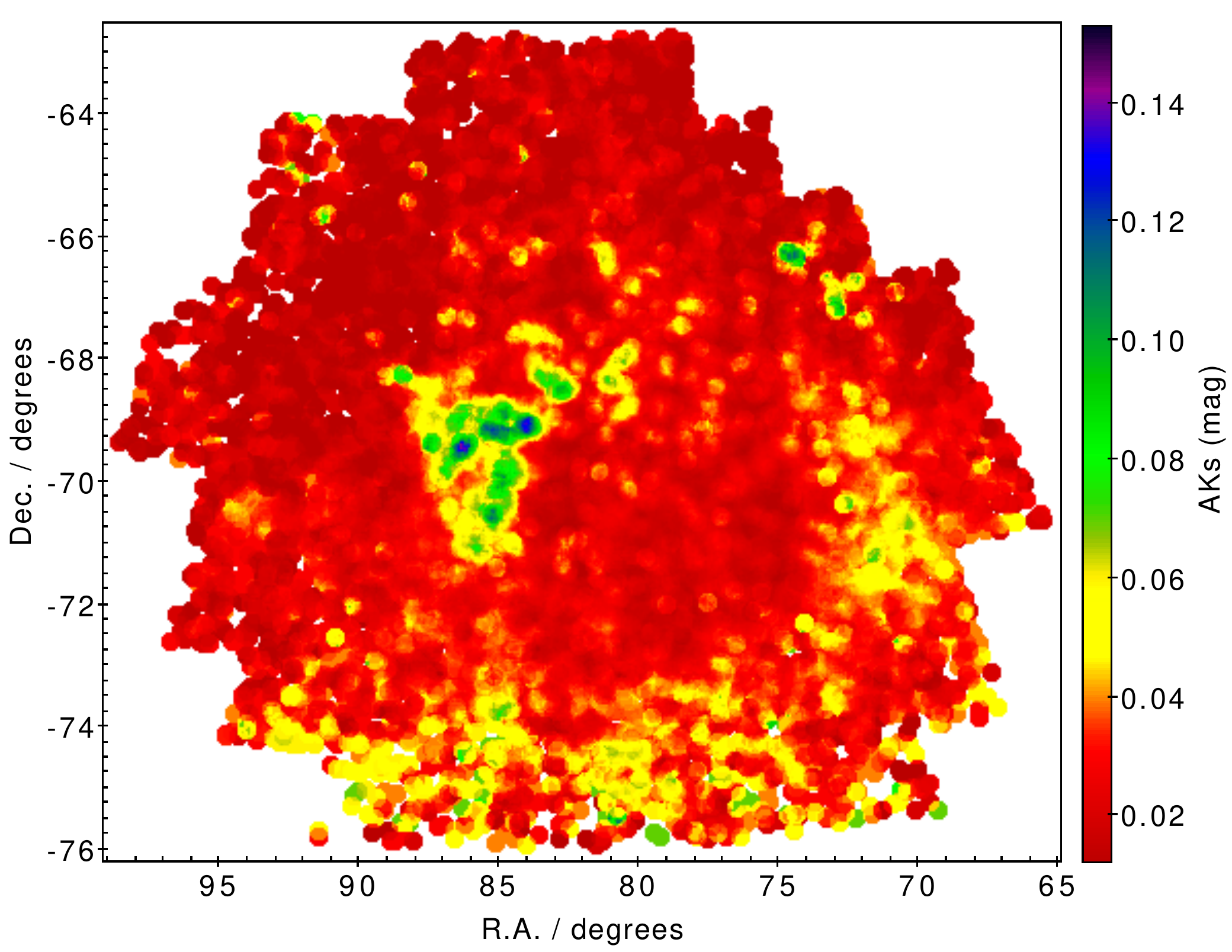}
\end{center}
\caption{{\it K$_\mathrm{s}$}  band absorption map derived from the RRab stars. The position of each RRL was re-binned  to a circle of 30 arcsec in diameter.}
 \label{fig:redmap}
\end{figure}

\begin{figure}
\begin{center}
\includegraphics[width=8.9cm, height=14cm]{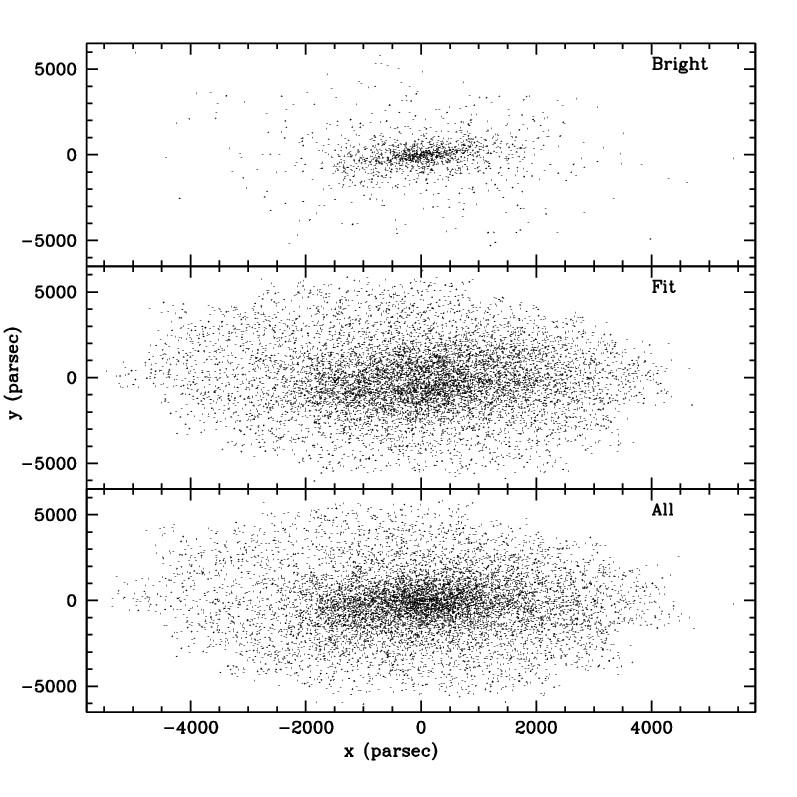}
\end{center}
\caption{{\it Bottom:} Spatial distribution of the 
whole sample of RRLs ($\sim 29,000$) considered in the paper;  {\it Middle:}  distribution of the RRLs which were actually used to fit  the  $PL_{K_\mathrm{s}}$ relation ($\sim 22,000$); {\it Top:}  distribution of the RRLs which appear to be overluminous in the $PL_{K_\mathrm{s}}$ relation ($\sim 7,000$), hence were discarded. They are mainly located along the LMC  bar, thus explaining the gap seen in that region in the middle panel.}
 \label{fig:trip}
\end{figure} 

\subsection{Reddening from  the RR Lyrae stars }

For each RRL pulsating in the fundamental mode a reddening estimate was obtained using  the relation derived by \cite{Pie02} which provides 
the intrinsic $V-I$ colour of fundamental-mode RRLs as a function of the star's period and amplitude in the $V$ band:
\begin{eqnarray}\label{eq:Piersimoni02}
\nonumber{(V-I)_0 = 0.65(\pm0.02) - 0.07(\pm0.01){\rm AmpV}+} \\
{+ 0.36(\pm0.06)\log P ~~~~~~\sigma=0.02}. 
\end{eqnarray}
where periods and amplitudes (in the $I$-band)  for the RRLs were taken from the 
OGLE~IV catalogue  \citep{Soszynski2016}, and the $I$-band amplitudes were transformed to
$V$-band amplitudes by adopting a fixed amplitude scaling factor of AmpV$/$AmpI=1.58. This value was 
derived from a statistically significant (75) data set of RRLs  in Galactic GCs (see \citealt{DiC11}). The intrinsic $V-I$ colours were then used to derive individual $E(V-I)$ reddening values  for each RRL. These 
 $E(V-I)$ values were then converted to extinction in the $Y$, $J$ and $K_\mathrm{s}$ passbands using the  coefficients  : $A_Y$ = 0.385A$_V$, $A_J$ =
0.283$A_V$, and $A_ K$ = 0.114A$_V$ with $A_V$= 2.4$E(V-I)$
 from 
\citet{kerber2009}, which were derived from the \citet{cardelli1989} extinction curve.
The  RRL absorption map in the {\it {K$_\mathrm{s}$}} band derived from the RRab stars  is shown in Figure~\ref{fig:redmap}. This map was used to estimate by interpolation the absorption for the RRLs in the 
$Gaia$ sample which lack $V$ and $I$  magnitudes and for the RRc stars 
for which Equation \ref{eq:Piersimoni02} is not applicable.   The complete
set of   $A_{K_\mathrm{s}}$ values is given in the ninth column of table~\ref{tab:RR88}.
It is interesting to note how
 the most active star forming regions in the LMC like
  30 Doradus pop up in the map. All over the LMC the average extinction is small and 
 its value in the $K_\mathrm{s}$  band  is: $A_{K_\mathrm{s}}$=0.03 mag with $\sigma=0.02$ mag.
A comparison of the reddening values derived
from RC stars  \citep{Tat13} and 
the values estimated from RRLs using the \cite{Pie02} formula shows that there is a  qualitative
agreement between the  two methods.  However, this agreement should be taken with some caution as \citet{Tat13}'s work is based on stars  
in a limited area of the LMC,  surrounding the 30 Doradus region. 
 Indeed, substantial differences between  the reddening maps
derived from RCs, background galaxies and RRLs were found for example in the SMC by \citet{bell2020}. These differences most likely 
arise from RCs, RRLs and background galaxies sampling regions at different depths.  In fact, RCs are expected to be embedded in the dust layers, RRLs probably are half in front and half in the background, and galaxies are definitely in the far field. One can thus expect some spatial correlations in the mean extinction values, but not identical values from these different indicators.

\subsection{Blended sources}\label{sec:blends}
Our first attempt to derive the $PL_{K_\mathrm{s}}$ relation  
has revealed a number 
of RRLs brighter than the main relation. 
We investigated their spatial distribution and found that 
they are mainly located in the central part of the LMC.  This is shown in  Figure~\ref{fig:trip} which presents in the bottom panel the spatial distribution of  the whole sample of RRLs considered in this paper ($\sim 29,000$), in the top panel the distribution of the RRLs which appear to be overluminous in the $PL_{K_\mathrm{s}}$ relation and, in the middle panel,  the distribution of the  RRLs  which were actually used to fit our final $PL_{K_\mathrm{s}}$ relation.
Further investigations, performed using the Fourier parameters ($\phi_{31}, \phi_{21}, R_{21}$ and $R_{31}$)
of the light curves  available in the OGLE~IV catalogue did not show any particular properties of the  overluminous RRLs. On the other hand, in the period--amplitude diagram based on  the $I$ amplitudes available in the OGLE~IV catalogue, the bright RRLs all show small, in some cases near-zero amplitudes, compared with  regular RRLs
of the same period. The decrease in amplitude at a given period can be  owing to these RRLs being blended with  non-variable stars.
We expect the centroid of a blended source to be determined  with poor accuracy
\citep[see e.g.][]{Rip14,Rip15}. For this reason  as a further test we plotted the distribution of distances in arcsec of the VMC sources   cross-matched with  the OGLE~IV RRLs. 
A clear separation is now seen in the two samples, for 94\% of the RRLs lying 
on the $PL/PW$ relations the cross-match radius is less than 0.2 arcsec, whereas for 68\% of the overluminous RRLs the 
cross-match radius is larger than 0.2 arcsec.
The average  accuracy of the VMC astrometry is on the order of 0.080 arcsec both in R.A. and in Dec. 
\citep{Cio11}. 
We therefore 
discarded the RRLs with a cross-match radius larger than 0.2 arcsec. 
A total of 3252 objects were discarded. This procedure allowed us to significantly reduce the scatter on the $PL/PW$ relations.
We  visually inspected the VMC images of some of the 
discarded RRLs, confirming that they all are clearly blended with stars and/or background galaxies.
Similar investigations were performed in the past and the same effects were noted, e.g. by
\citet{Rip15} for Type II Cepheids.
The final sample of  LMC RRLs  after this cleaning procedure contains  25,795 objects.   This is the sample that was used as a starting point to investigate the $PL$ and $PW$
relations presented in the following sections. Additional RRLs were later discarded from the final 
fit 
based on a 3-$\sigma$ clipping procedure leading to 
$PL_{K_\mathrm{s}}$ relations using a clean sample of $\sim 22,000$ RRLs.

 \subsection{Metallicity of the RR Lyrae stars}\label{subsec:RRmet}
Knowledge of the metallicity ($Z$) is needed 
 to construct 
$PLZ$ and $PWZ$ relations for RRLs. 
However, spectroscopic metallicities are available  only for a very limited number of LMC RRLs (e.g. \citealt{Gra04,Bor04,Bor06} and references therein).  
\cite{JK96} 
and \cite{Mor07} showed that it is possible to derive a ``photometric" estimate of the 
metal abundance ([Fe/H]) of  an RRL of known pulsation  period from the $\phi_{31}$ Fourier parameter  of the 
 $V$-band  light curve decomposition.
 A new calibration of the $\phi_{31}$--[Fe/H]  relation  for fundamental-mode  and 
 first-overtone
RRLs was  published by \cite{Nem13}, 
 based on excellent 
accuracy RRL light curves obtained with the {\it Kepler} space telescope,  
and metallicities derived from high resolution spectroscopy. 
This new calibration provides metal abundances 
directly tied to the metallicity scale of  \cite{Car09} which is based on high dispersion spectroscopy  and holds 
for the metallicity range from [Fe/H]$\sim$ 0.0 to  $\sim -$2.6 dex.  

\citet{Skowron2016} applied the calibration of  \cite{Nem13} to the OGLE-IV fundamental mode 
RRLs, obtaining a median metallicity value  of  [Fe/H]=$ - 1.59 \pm 0.31$ dex for the 
LMC on the \citet{zinwe1984} scale. Metallicities from \citet{Skowron2016} are available for 16,570  LMC RRab stars in our sample. This sample was used to compute a metallicity map
of the LMC.  Through the interpolation of this map  it was possible  to give a metallicity  estimate  also  to the RRc stars.
The single  [Fe/H] values in \citet{Skowron2016} and the ones derived from the 
metallicity map, 
were used to compute the metallicity-dependent relations listed in Tables~\ref{tab:pl} and \ref{tab:pw}.
A non-linear least-squares Marquardt--Levenberg  algorithm (e.g. \citealt{Mighell1999}) was used to fit the $PLZ$ and $PWZ$ relations along with a 3 $\sigma$ clipping  procedure to clean from outliers. Our 
final $PLZ$ and $PWZ$ relations are based on 
 a sample $\sim 13,000$ RRab  and $\sim 5,000$ RRc stars
(see Tables~\ref{tab:pl} and \ref{tab:pw}).

\subsection{Derivation of the $PL(Z)$ and $PW(Z)$ relations}

\begin{figure}
\includegraphics[width=8.9cm, height=9.9cm]{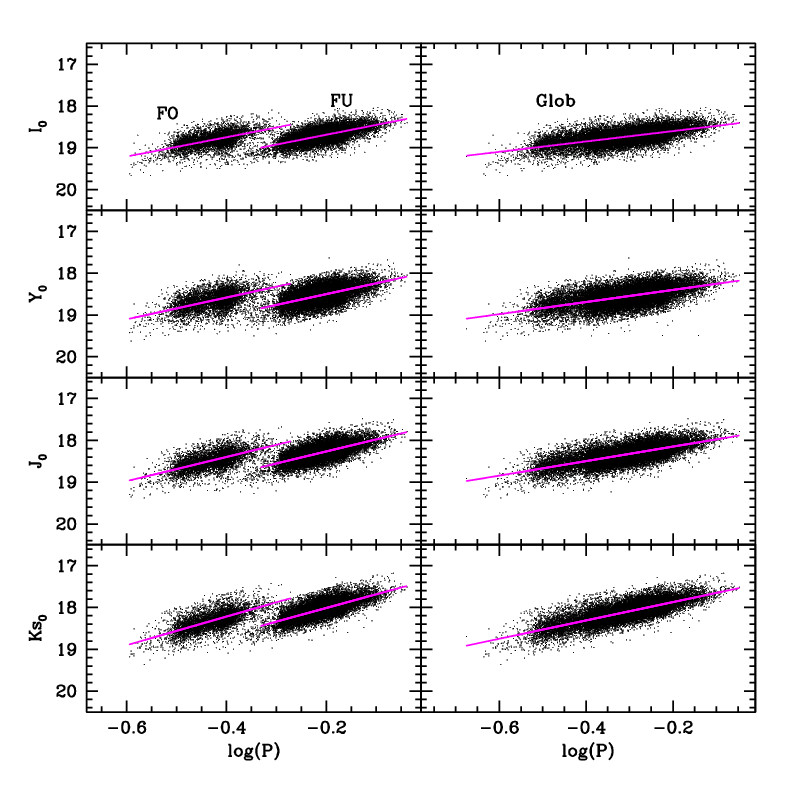}
\caption{ De-reddened mean  magnitudes versus $\log P$, 
for RRLs in the LMC. Black points are stars retained for the final $PL$ computation. 
The left-hand panels show RRc (first overtone, FO) and RRab stars (fundamental mode, FU) separately. The right-hand panels show all RRLs after  the $\log P$ of the RRc stars was fundamentalised. The  magenta lines show the $PL$ relations derived using these data (see Table~\ref{tab:pl}).}
\label{fig:PLk_all}
\end{figure}

$PL_{K_{\mathrm{s}}}$(Z) relations were computed using the periods,   $\langle K_\mathrm{s} \rangle$ mean magnitudes, absorption values and metallicity for the RRLs listed in  Table~\ref{tab:RR88}.  We fitted a linear least-squares relation of the form: 
$m_{X0} = a + b\times\log P  + c\times$[Fe/H].
  For the global relation (Glob), which includes RRab plus RRc stars,
 the periods of the RRc stars were 
fundamentalised using the classical  relation $\log P_{FU}{\rm (RRc)}$ = $\log P_{\rm RRc}$ + 0.127 
\citep{Iben74}. 
Only stars with $E(V-I)>0$ mag were considered and an unweighted least-squares fit 
with  3 $\sigma$ clipping was applied, hence reducing the sample of RRLs actually used to 
construct the $PL$ relations to around 22,100 sources ($\sim$16,900 RRab and $\sim$5,200 RRc).  
The same procedure was used to obtain $PL$ relations in the $Y$ and $J$ bands. 
There are on average only  3--4 phase points for each RRL in the VMC data available in the $Y$ and $J$ bands   
(see Table~\ref{tab:lc}).
Hence, the resulting average magnitudes
have larger errors and 
the $PL$ 
relations in $Y$ and $J$ have larger rms values than for the $K_\mathrm{s}$ \ band. 
In the optical we obtained  a $PL$ relation in the $I$ band using the OGLE~IV data. 
 The $PL$ and $PLZ$ relations derived  using  this procedure are summarised in Table~\ref{tab:pl}. 
In the case of the   $PL_{K_\mathrm{s}}$  we also derived 
individual relations 
for each VMC tile and their parameters  are given in Table~\ref{tab:matching}.
Figure~\ref{fig:PLk_all} shows the  $PL$ relations in the $I$, $Y$, $J$ and ${K_\mathrm{s}}$   bands obtained  for RRab (FU), RRc (FO) on the left and  for RRab plus fundamentalised RRc stars (Glob) on the right. 

\begin{figure}
\includegraphics[width=8.9cm, height=10.1cm]{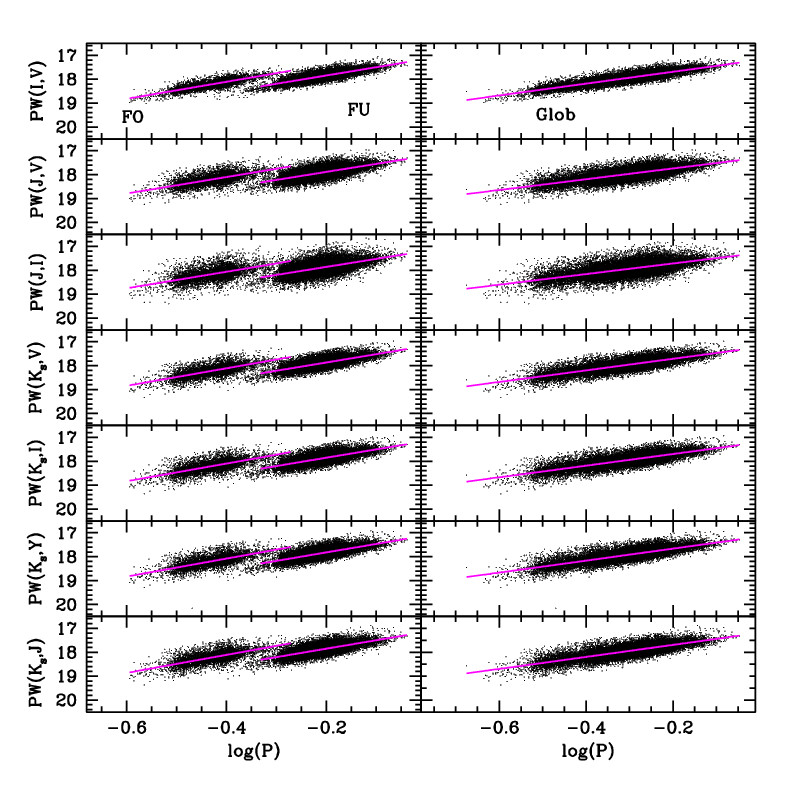}
\caption{ As in Figure \ref{fig:PLk_all} but for Wesenheit magnitudes.
 The  magenta lines show the $PW$  relations presented in Table~\ref{tab:pw}.}
\label{fig:pws}
\end{figure}

Reddening-independent $PW$ relations can be obtained by combining magnitudes and colours in different bands. These reddening free magnitudes  were introduced
by \cite{vandenB1975} and \cite{madore1982}. They are defined as: $W(X,Y)=m_X$ -- $R(m_Y$--$m_X$) where $X$ and $Y$ are two different passbands and $R$ is the ratio between
selective absorption in the $X$ band and colour excess in the adopted colour. These coefficients are fixed according to  the \cite{cardelli1989} law.  
  A set of $PW$ and $PWZ$ relations was obtained fitting the VMC $Y$, $J$ and $K_\mathrm{s}$  and OGLE~IV $V$, $I$  photometry  to the equation $m_X - R\times(m_Y - m_X)= a+ b\times\log P  + c\times$[Fe/H]. The parameters obtained
are summarised  in Table~\ref{tab:pw}. 
 The PW relations in different bands are presented 
in Figure~\ref{fig:pws}.

\subsection{{\it Gaia}  Proper Motions for the LMC RR Lyrae stars}\label{gaia}
 The recently released {\it Gaia} eDR3 catalogue \citep{brown2020} includes  proper motions for 21,801 RRLs in our catalogue, 
with average errors of $\sigma_{\mu_{\alpha}^{*}} \sim 0.30$ and $\sigma_{\mu_{\delta}} \sim  0.32$ mas yr$^{-1}$, respectively.
We selected RRL stars in the LMC  retaining only
sources with proper motions within 1$\sigma$ from the average proper motions in R.A. and Dec. of the whole sample of 21,801 RRLs, 
that are  $\langle{\mu_{\alpha}^{*}}\rangle=1.85$ and $\langle{\mu_{\delta}}\rangle=0.36$ mas yr$^{-1}$, respectively. 
This resulted in a subsample of 4690 RRLs. 
A fit of the  PL$_{K\mathrm{s}}$ relation using only this {\it Gaia} proper motion-selected sample of RRLs provides:   $a$= 17.44 mag, $b=-$2.55 mag, $\sigma_a = 0.01$ mag, $\sigma_b = 0.03$ mag,   rms= 0.14. 
This is consistent with the global relation (Glob) presented in  Table~\ref{tab:pl}.
 A similar analysis was performed for the other PL(Z) and PW(Z) relations  presented
in table~\ref{tab:pl} and \ref{tab:pw} obtaining always consistent 
results. This gives  
us confidence that the $PL$ and $PW$ relations derived in this paper are based on samples of RRLs which are bona fide LMC members.

\subsection{Comparison with theory and other papers}
As a final step of our analysis of the RRL $PL$  relations
we have compared the $PLZ$ and $PWZ$ relations  obtained here with the theoretical results of  \citet{marconi2015}.

\begin{figure*}
\begin{center}
\includegraphics[width=8.7cm, height=4.3cm]{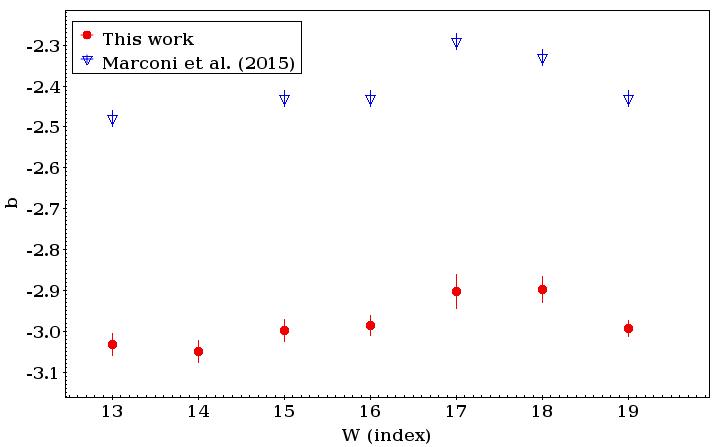}\includegraphics[width=8.7cm, height=4.3cm]{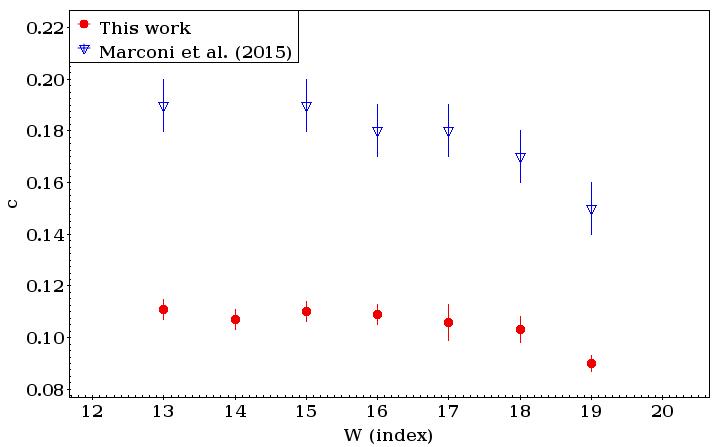}
\end{center}
\caption{  Comparison of the coefficients of the PWZ relations for RRab stars derived in this paper  with those of the corresponding  relations of \citet{marconi2015}. In the left-hand panel are shown the  period slopes, while 
 the right-hand panel shows the metallicty slopes. The $W$ indices are labelled as follows: $13=PW(K_{\mathrm{s}}, J),
14=PW(K_{\mathrm{s}}, Y), 15=PW(K_{\mathrm{s}}, I), 16=PW(K_{\mathrm{s}},V), 17=PW(J, I), 18=PW(J, V)$ and $19=PW(I, V)$.
}
 \label{fig:confr}
\end{figure*}

Three of our $PLZ$ and six of our $PWZ$  relations have a counterpart in \citet{marconi2015}. 
A comparison of the 
matching relations 
shows that the slopes in period and  metallicity are different,
with an average difference of 0.53($\pm0.05$) dex units in the $\log P$ term and 0.07($\pm0.01$) dex units in the metallicity term.  This is clearly seen in  Figure~\ref{fig:confr} where
the parameters of some PWZ relations derived in this work are compared with those in the corresponding  relations by \citet{marconi2015}.

\begin{figure*}
\begin{center}
\includegraphics[width=8.7cm, height=4.6cm]{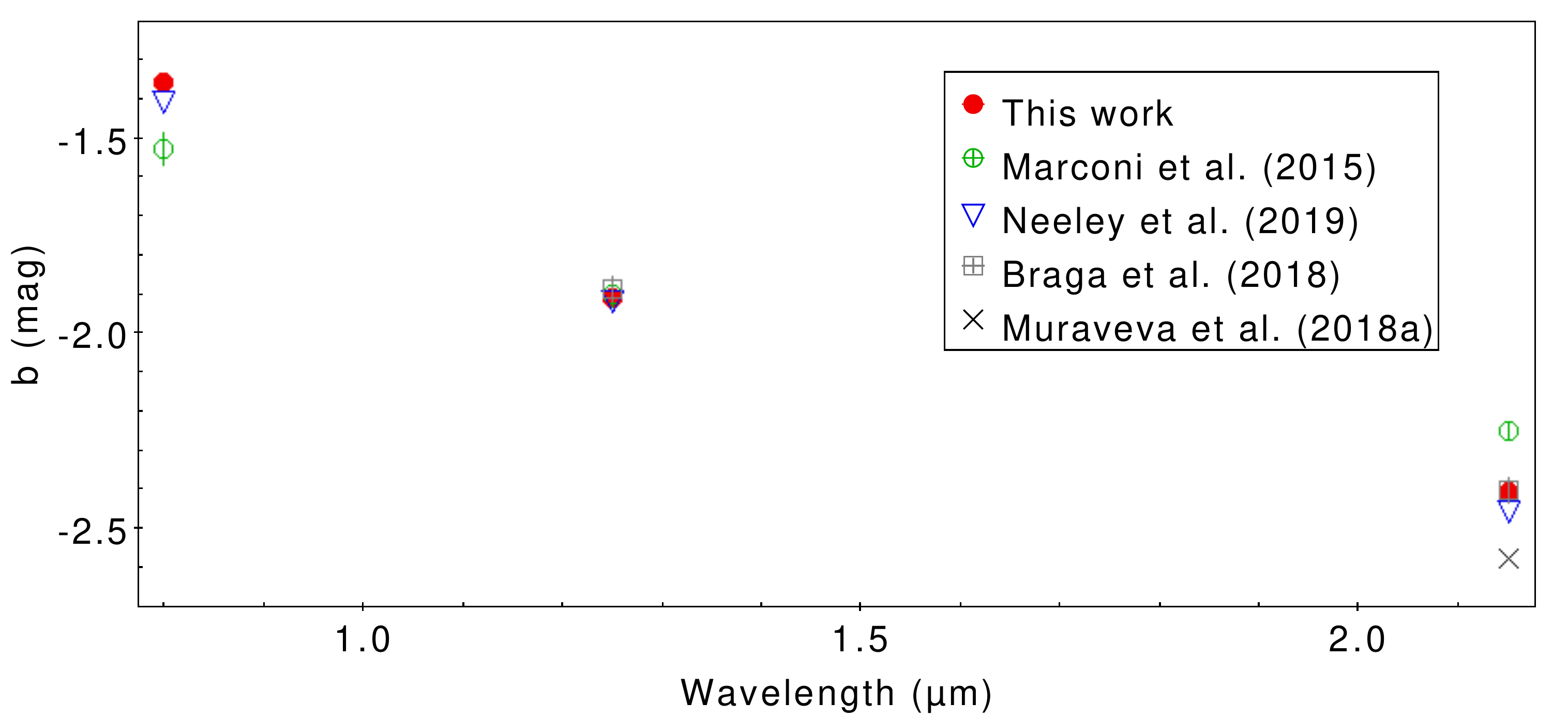}\includegraphics[width=8.7cm, height=4.6cm]{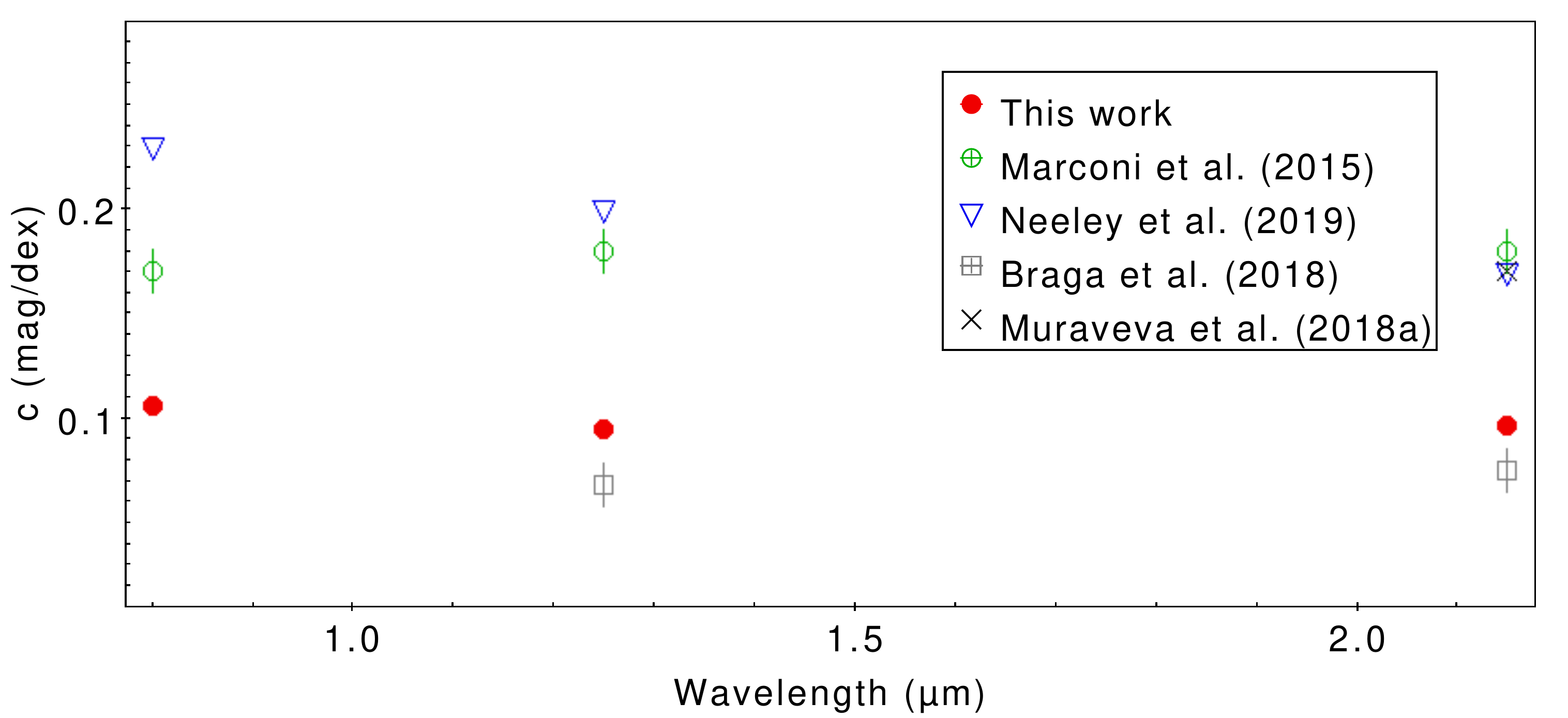}
\end{center}
\caption{Comparison of the coefficients of the PLZ relations for RRab + fundamentalised RRc stars derived in this paper  with those indicated in the legend. 
The period slopes are shown in the left-hand panel and the metallicity slopes in the 
right-hand panel.
The errors of Neeley et al. (2019) and Muraveva et al. (2018a) are not reported 
to avoid confusion. 
}
 \label{fig:confrpl}
\end{figure*}

These differences most likely arise from the larger dependence on metallicity of  the theoretical relations,  which predict metallicity terms on the order of 0.2 mag dex$^{-1}$. 
Further comparisons with the theoretical pulsation scenario will be discussed in a future paper (M.~I.~Moretti et al., in preparation).
Such a large metallicity dependence is not observed for several  empirical RRL $PLZ$ relations in the literature \citep[e.g.][]{Del06, Sol08, Bor09}.
However, using literature data for Milky Way (MW) field RRLs,  \citet{Mur18a} derived an empirical absolute $PLZ(K)$ relation calibrated using {\it Gaia} DR2 parallaxes, which shows a non-negligible dependence on  metallicity,  in the same direction as that found by the theoretical $PLZ$ relations.  Similarly, \citet{sesar2017}  and then  \citet{neeley2019} using {\it Gaia} DR2 parallaxes found for the metallicity slopes of the PLZ relations values between 0.17 and 0.20 mag dex$^{-1}$. 
A possible explanation for the differences in the metallicity term (and, in turn, also for  the $\log P$ term)
 is that the RRLs considered by the   \citet{marconi2015}, \citet{Mur18a}, 
 \citet{neeley2019} and \citet{sesar2017}  span a range in metallicity from $Z$ = 0.0001  to $Z$ = 0.02, while the metallicity  of the LMC RRLs covers a much smaller range  from $Z$=0.0001 to $Z$=0.001 \citep{carrera2008}, hence lacking the more metal-rich component observed in the MW.
 A comparison between the  coefficients of the PLZ relations for RRab + fundamentalised RRc stars derived in this paper and in those cited above is shown in Figure~\ref{fig:confrpl}. 
 The fit parameters differences range from  an almost consistency in 
 1 $\sigma$ in the $J$ band b parameters to  a 6 $\sigma$ difference for c parameters in the same band.
 Consistently larger differences arise in the $Y$,$J$ and $K_{\mathrm{s}}$ bands for the 
 c parameters. While for the b parameters the largest difference are seen especially for the
 $K_{\mathrm{s}}$ band. 

\section{Structure of the LMC}

\subsection{Individual distances to RR Lyrae stars}

The $PL$ and $PW$ relations derived in the previous sections can be used
to infer the distance to each RRL in our LMC sample
once a zero point is adopted.
 For this analysis we do not rely on the PLZ relations 
as they are based on a smaller sample than  the PL relationships.
We assumed as distance to the LMC centre the estimate 
 of \citet{pietrink2013} which corresponds
to $d_0= 49.97 \pm 1.12$ kpc or to a distance modulus of $(m-M)_0$ =18.49 mag.
However, this choice does not affect our study of the LMC's structure, for which we rely on differential distances.

We measured the distance to each RRL using the  $PL_{K_{\mathrm{s}}}$ 
relation and 
two different versions of the $PWZ$s. 
Results obtained from  the three different relations were then compared. 
Since the $PWZ$s include a metallicity term, their application is limited  to the RRab and RRc sample for which we have metallicities.
We chose to adopt the  $K_\mathrm{s}$, $J-K_\mathrm{s}$ and the $K_\mathrm{s}$, $V-K_\mathrm{s}$ $PWZ$
 relations. We adopted the former  
because the $K_\mathrm{s}$ and $J$ bands are less sensitive to possible 
deviations from the colour correction,  
 and the latter because 
it is 
weakly dependent on the colour term ($R$=0.13). 
We opted for the $PL_{K_{\mathrm{s}}}$  relation
because it is less affected by reddening, and used the global version, which allowed us to determine the distances  to 22,122 RRLs  of both RRab and RRc types. The average
error in the distance estimate for each RRL is of the order of $\sim 3.5 $ kpc ( $\sim 0.15$ mag).
The mean difference between the distances measured using the two $PWZ$ relations is 0.45 pc, whereas the mean difference between distances from the $PW(K_\mathrm{s}$, $V-K_\mathrm{s}$) and the  $PL_{K_{\mathrm{s}}}$ relations is 23 pc ($\sim 0.001$ mag), which is very small compared with the average individual  
error of each RRL.
In the following analysis we rely on the distances derived using the $PL_{K_{\mathrm{s}}}$ 
since  it is based on a larger sample of RRLs, it has the smaller rms 
between the PL relations derived here and the ${K_{\mathrm{s}}}$ band is the one that is less effect 
by reddening.

\subsection{3-D geometry of the LMC}
Once the distance to each RRL is determined, 
it is possible to derive the Cartesian distribution of the RRLs from the R.A., Dec coordinates and the individual  distances using the method described by  \citet{VC01} and \citet{wei2001}:

\begin{align}
x  &=  -d~{\rm sin}(\alpha - \alpha_0){\rm cos} \delta \nonumber \\
y  &= d~{\rm sin} \delta {\rm cos} \delta_0 - d~{\rm sin} \delta_0 {\rm cos} (\alpha - \alpha_0 ) {\rm cos} \delta \\
z  &= d_0 - d~{\rm sin}\delta {\rm sin} \delta_0 - d~{\rm cos} \delta_0 {\rm cos} (\alpha - \alpha_0) {\rm cos} \delta \nonumber ,
\end{align}
where $d$ is the individual distance to each RRL, 
$d_0$ is the distance to the LMC centre 
and ($\alpha_0$, $\delta_0$) are the R.A. and Dec. coordinates of the LMC centre.
We assume as reference system ($x,y,z$) one that has the origin
in the LMC centre at ($\alpha_0, \delta_0$ and $d_0$),
 the $x$-axis anti-parallel  to the R.A.-axis, the $y$-axis 
 parallel to the Dec. axis and the $z$-axis pointing towards the observer.
The $\alpha_0$, $\delta_0$  coordinates of the LMC centre  were derived from the average
position of the RRLs  ($\alpha_0 = 80^\circ.6147 ,\delta_0=-69^\circ.5799$). 
Our centre is offset by  $\sim 0.8^\circ $ from the LMC centre 
 of  \citet[][$\alpha_0 = 81^\circ.24 ,\delta_0=-69^\circ.73$]{Youssoufi2019},  which was 
obtained using stellar density maps of all the stellar populations in the LMC. 

\begin{figure}
\begin{center}
\includegraphics[width=8.9cm, height=12cm]{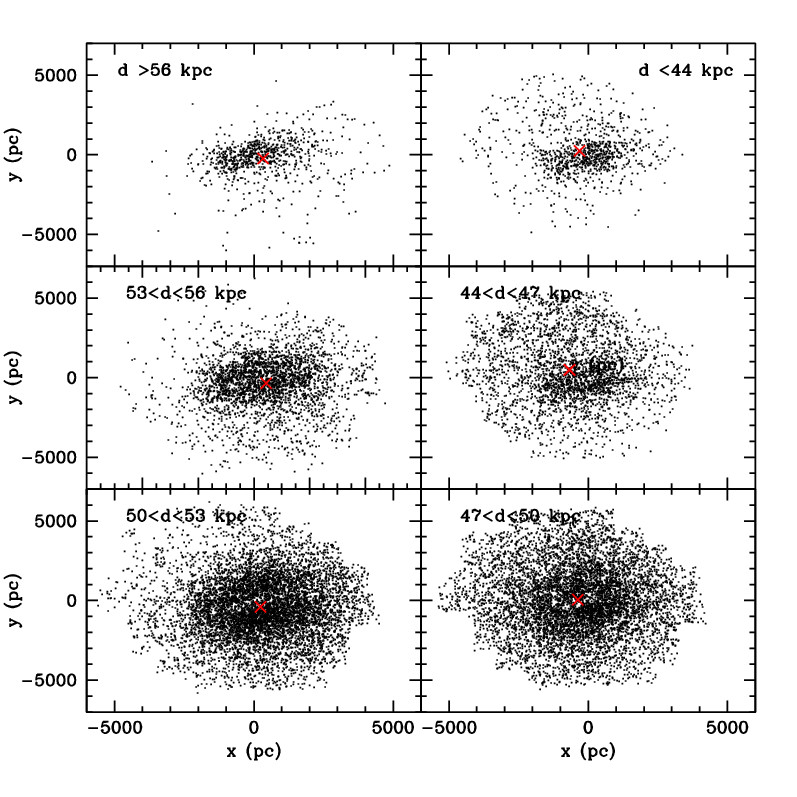}
\end{center}
\caption{LMC RRLs divided into bins of heliocentric distance, as indicated by the labels. The average centre of the distribution in each panel is marked with a red cross.}
 \label{fig:slice}
\end{figure}

As a first approach we divided the LMC into ($x,y$) planes by considering different intervals in distance.
Figure~\ref{fig:slice} shows the distribution of the LMC RRLs in Cartesian coordinates divided 
into bins of distance.   The bins were  of the same order as the average distance standard deviation,   $\sim 3.5$ kpc. 
Going clockwise from the top right-hand panel the RRLs are mapped for increasing distance values.
A quick look shows that there are two extreme cases, with 
the RRLs closest (top right-hand panel) and most distant to us  (top left-hand panel) showing  different spatial distributions.
A fraction of the RRLs in the top right-hand panel  of Figure~\ref{fig:slice} appear to be spread all over the field and are likely RRLs belonging to the MW halo. 
This view  is corroborated  by the proper motions of the RRLs in this 
sample of which 43\% are outside of 1 $\sigma$ the average $\mu_{\alpha}^{*}$ and $\mu_{\delta}$
values.  These stars are placed mainly in the central region of this plot.
On the other hand,  it is also possible that this sample of RRLs belong to the 
outer most region of the elongated LMC halo, which is distorted by tidal forces owing 
to the interaction with the MW. This would explain why the proper motions are different from the average values in the main body of the LMC.  In the same direction  El Youssoufi et al. (submitted, see sect. 2.3) finds that the centre of the stellar proper motions for stars in the outer regions of the SMC is shifted from that of the inner regions, suggesting that these stars are probably associated to the LMC. 
In  the same sample a concentration of RRLs  likely belonging to the LMC
shows up right and south of the centre.
That is the part of the 
LMC which points towards the MW. 
The average coordinates of the RRLs in this panel  are $\langle (x,y) \rangle =(-309, 241)$ pc  and they represent  5.1\% of the whole sample.
In the middle right-hand panel are shown RRLs
with distances between 44 and 47 kpc and it is  still
possible to notice a protrusion of stars in the same region as in the previous panel. The centre coordinates  in this case  are $\langle (x,y) \rangle =(-667, 491)$ pc, for 12.5\% of RRLs.
Between 47 and 53 kpc (bottom two panels) there are RRLs that belong with
good confidence to the LMC and its halo. In the bottom right-hand  panel (47 $< d <$ 50 kpc)
the RRLs are distributed in what is reminiscent of a regular,  spherical shape and no particular structures are seen. The centre of the distribution is   at $\langle (x,y)\rangle =(-374,  51)$ pc
 for 31.2\% of the RRLs. 
In the bottom left-hand panel ($50< d <53$ kpc)  a lack
of RRLs along the central part of the LMC bar 
is visible  (see also the middle panel of  Figure~\ref{fig:trip}). This is the region of the highest crowding where many RRLs 
although recovered 
in the VMC catalogue, are found to be located above  the 
$PL$ relations owing to  contamination by close sources, and which hence were discarded (see Sect.~\ref{sec:blends}). 
A similar feature was detected using the OGLE IV catalogue by \citet{Jacyszyn2017}, who  likewise concluded that it  
is probably caused by  source crowding and blending.
The distribution peaks  at $\langle (x,y)\rangle = (241,  -401)$ pc  for 34.1\%  of the total sample.
In the middle left-hand panel there are stars with distances  between $53< d <56$ kpc, they represent  
 12.5\% of the total with centre coordinates  at $\langle (x,y)\rangle = (435, -343)$ pc.
The top left-hand panel shows the most distant RRLs, corresponding to  4.6\% of the total
sample. Their distribution peaks at $\langle (x,y)\rangle = (338,  -234)$ pc.
We thus conclude that the LMC has a regular shape
similar to an ellipsoid with  the north-eastern stars closer to us than  the  south-western component. 

No particular structures other than that of an ellipsoid are seen. 
To derive the parameters of the ellipsoid we used a similar approach as that used 
by \citet{debandsingh2014} and which is described in section 5.2 of their paper. 
The three axes of the ellipsoid that we derived are  $S_1$=6.5 kpc, $S_2$=4.6 kpc  and $S_3$=3.7 kpc.
We found an inclination relative to the plane of the sky and a position angle of the line of nodes (measured from north to east) of  $i=22\pm4^{\circ}$ and  $\theta=167\pm7^{\circ}$, 
respectively.  These results are in agreement with \citet{debandsingh2014} 
who found $i=24.02 ^{\circ}$ and  $\theta=176.01 ^{\circ}$. Our  result for
the inclination  also compares well  with other papers in the literature:  
 for example,  \citet{nikoalev2003} found  $i=30.7\pm1.1^{\circ}$  using CCs. The small difference in the values can
be attributed to the different stellar populations used as RRLs trace old stars which should be more smoothly distributed, while young stars such Cepheids are found in star forming
regions located mainly in the disc of the galaxy.

\begin{figure*}
\begin{center}
\includegraphics[width=5.9cm, height=5.9cm]{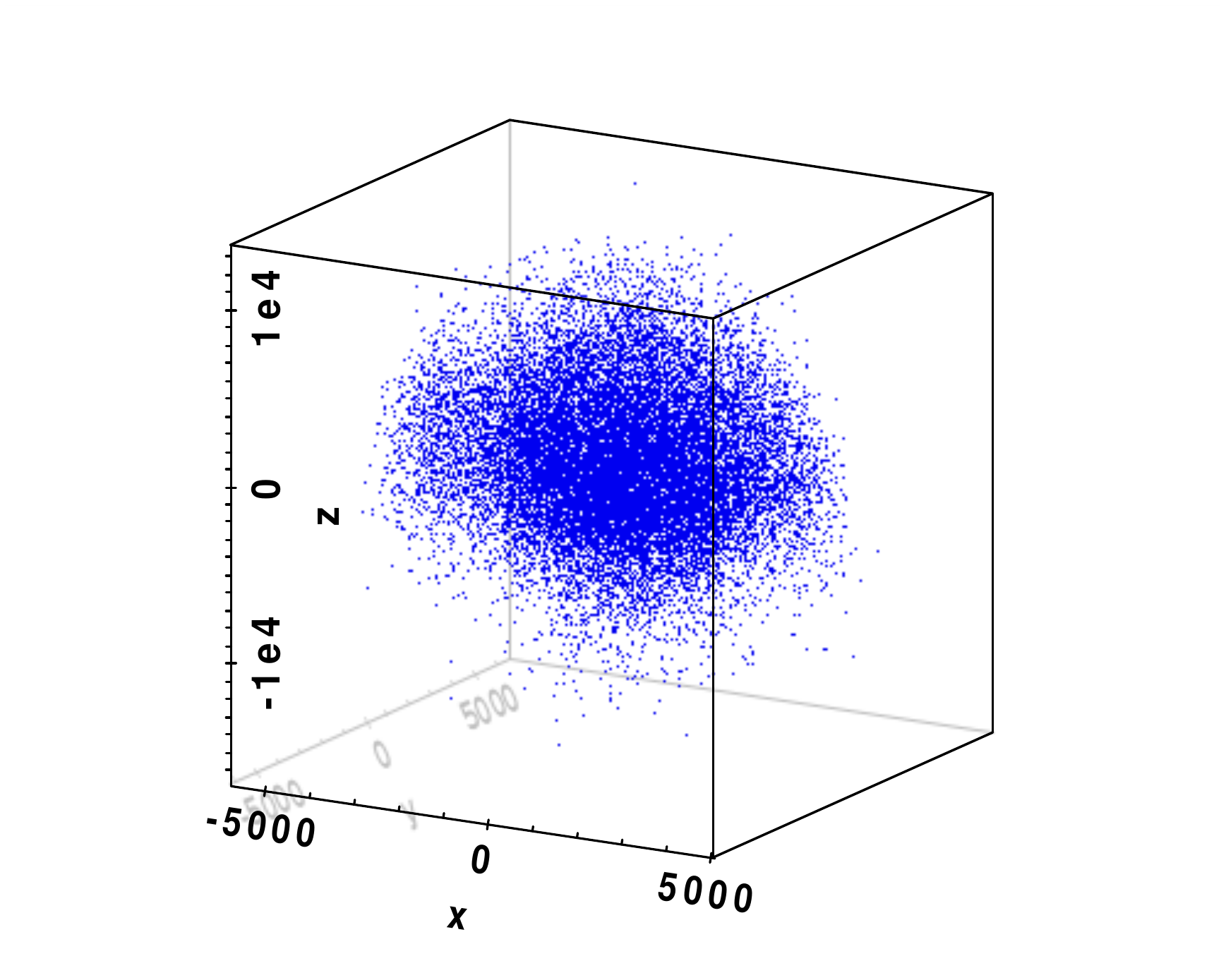}\includegraphics[width=5.9cm, height=5.9cm]{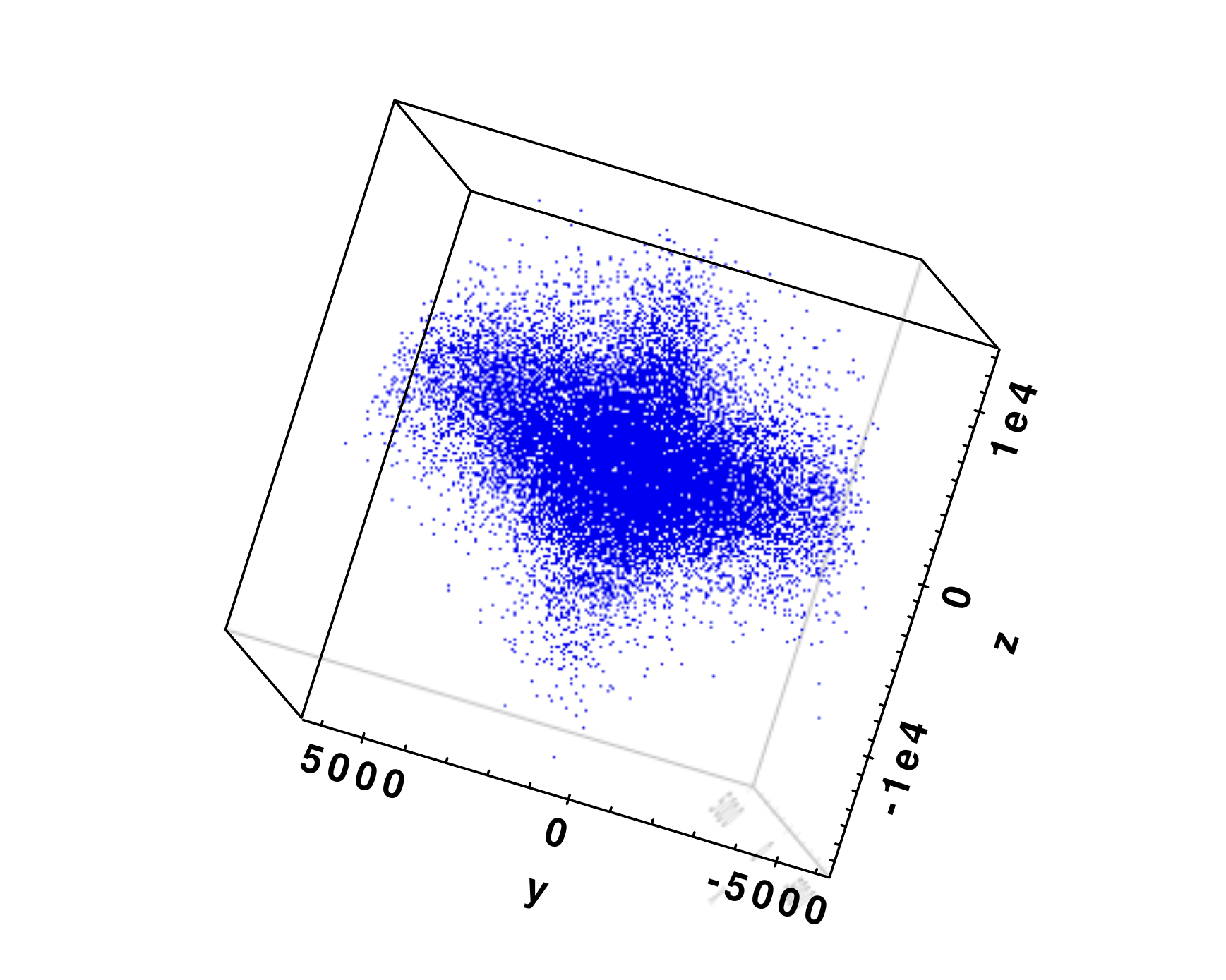}
\includegraphics[width=5.9cm, height=5.9cm]{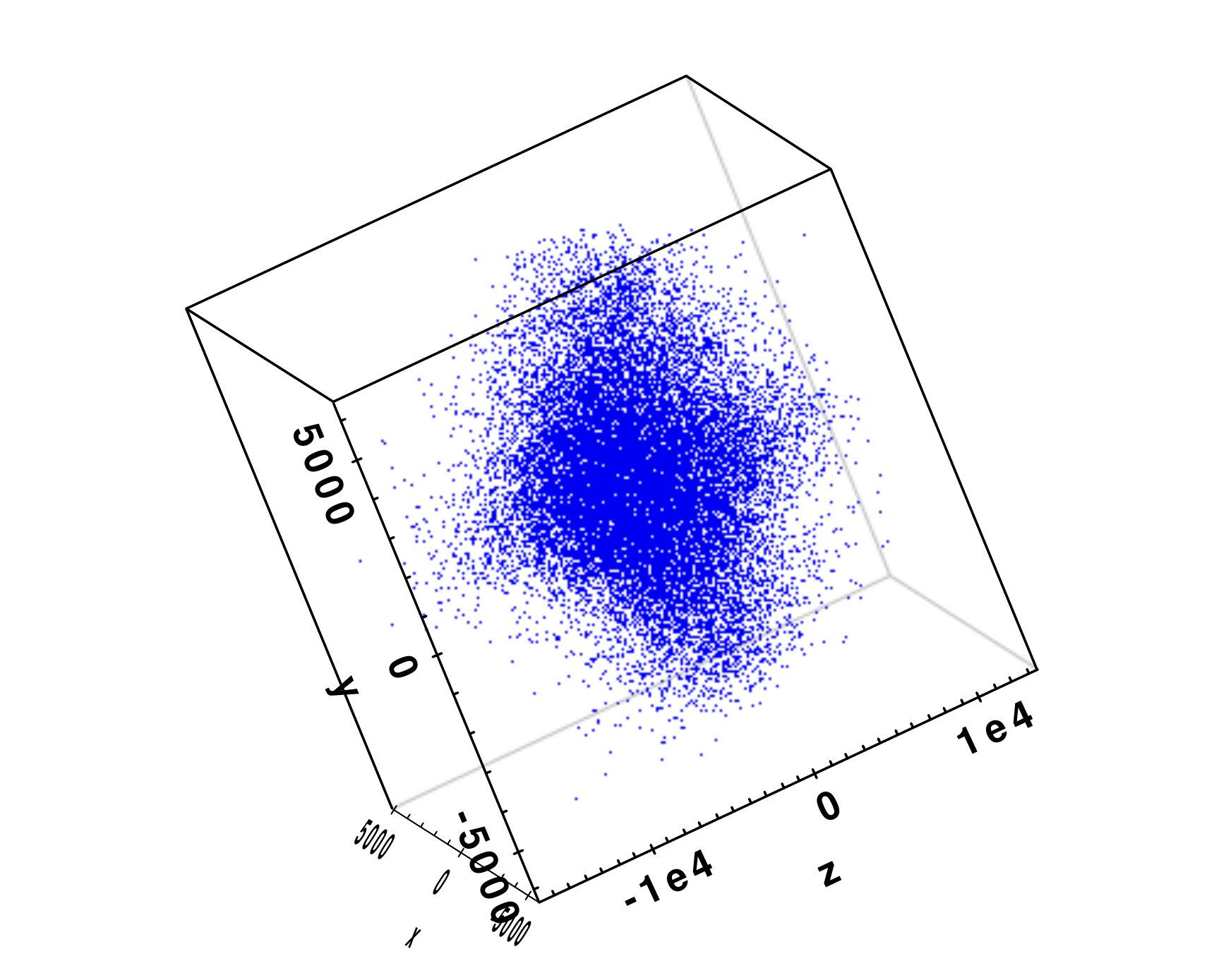}\includegraphics[width=5.9cm, height=5.9cm]{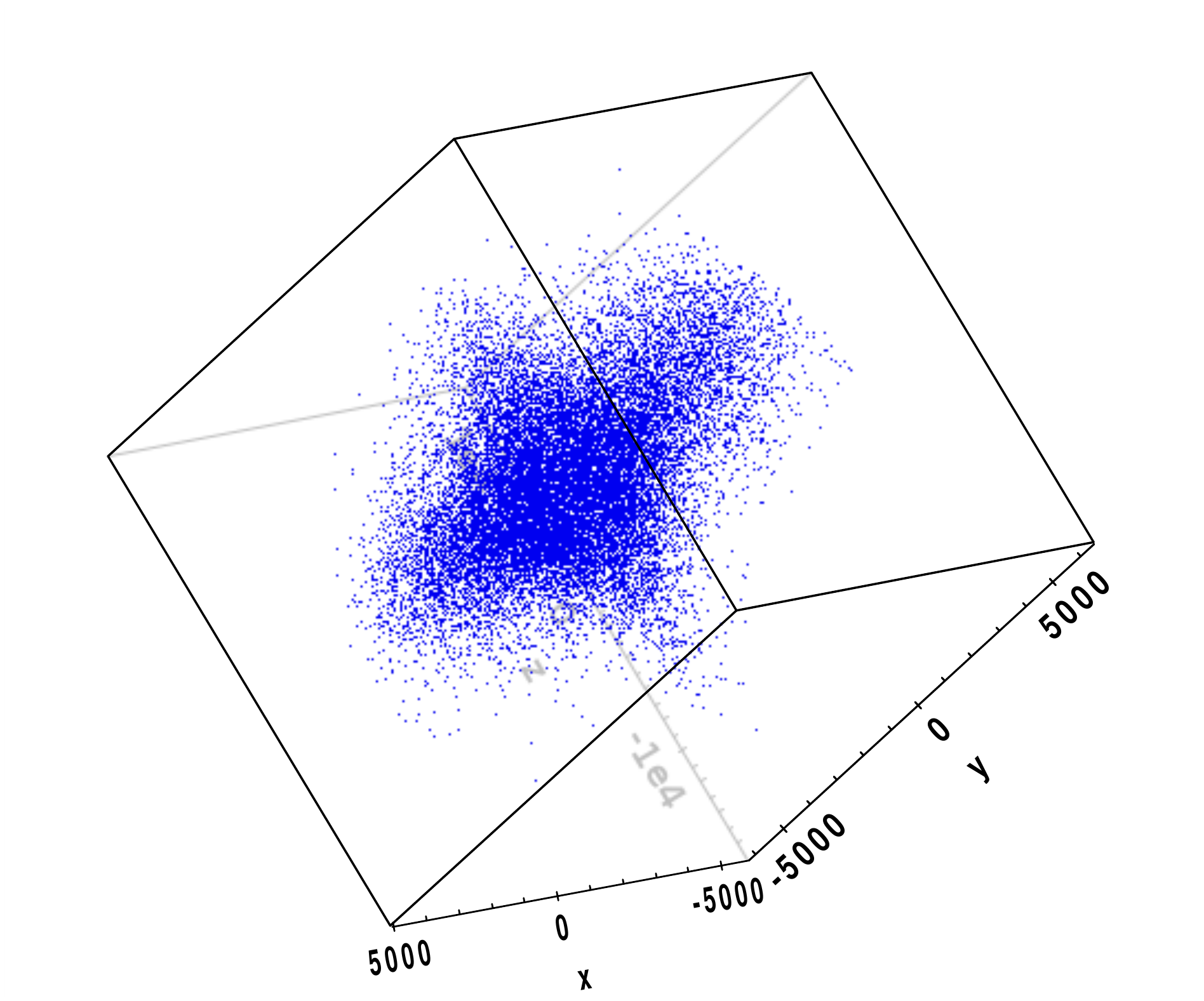}
\includegraphics[width=5.9cm, height=5.9cm]{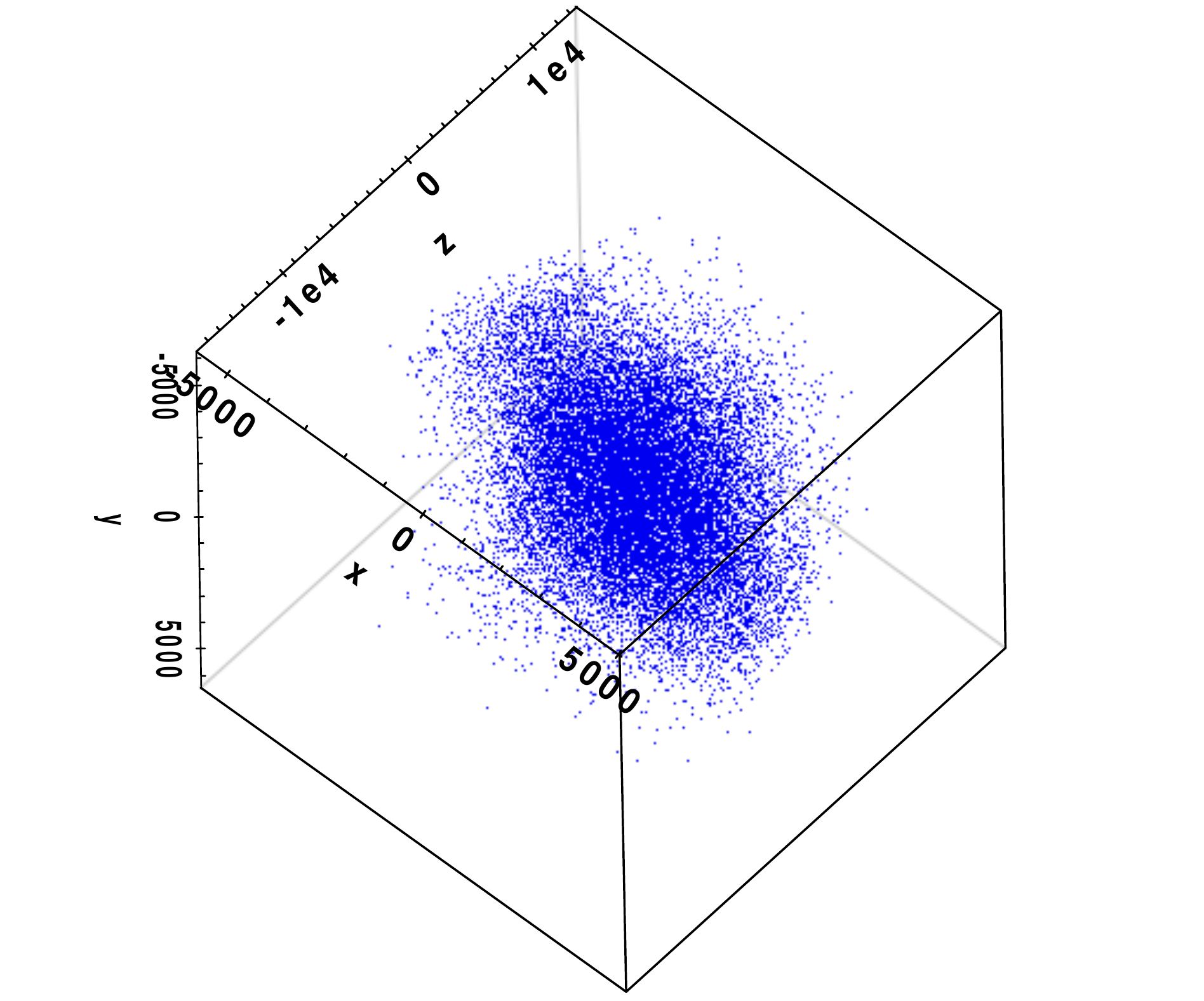}\includegraphics[width=5.9cm, height=5.9cm]{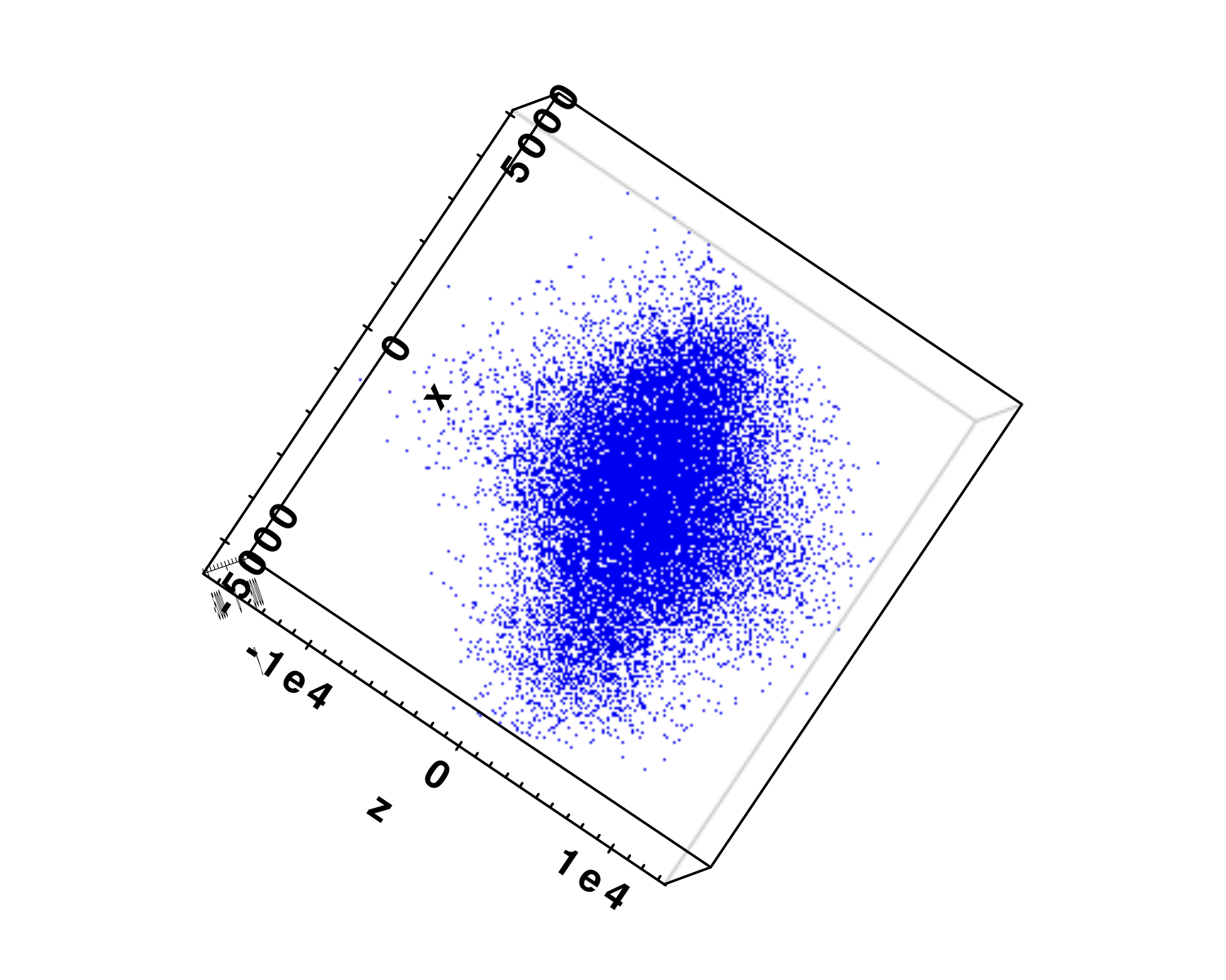}
\end{center}
\caption{3-D shape of the LMC  inferred from the RRLs, as seen from different viewing angles. 
The protrusion of sources extending from the LMC centre could be due to crowding effects.
Units along the axes are parsecs. 
}
\label{fig:3dshape}
\end{figure*}

Combining the results described above we can now visualize the 3-D shape of the LMC as traced by its RRLs.
This is shown in Figure~\ref{fig:3dshape} for different viewing angles. The overall LMC structure is that
of  an elongated ellipsoid with a protrusion of stars  more distant and closer to us extending from the centre of the LMC. This singular feature,  also found by other studies based on OGLE~IV RRLs 
\citep[see e.g.][]{Jacyszyn2017}, is not a real physical structure and is mainly due to blended  sources in the centre  of the LMC.  However we can not completely 
exclude that the innermost regions, which are at present significantly influenced by crowding, may harbour  additional substructures.
The regions of the ellipsoid extending to  the north-eastern  direction  are closer to us than those in the opposite  south-western  direction.
The difference in heliocentric distance between these two regions 
is on  the order of ~2 kpc. 

\begin{figure}
\begin{center}
\includegraphics[width=8.9cm, height=12cm]{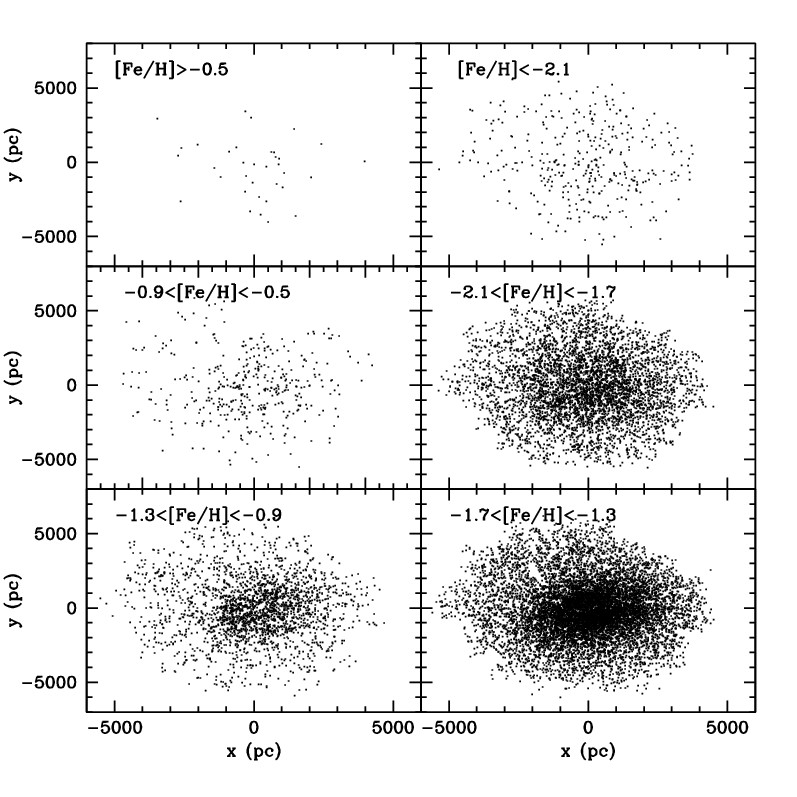}
\end{center}
\caption{LMC RRLs divided into bins of metallicity, as indicated by the labels.
}
\label{fig:met}
\end{figure}

The analysis was repeated by slicing  the LMC into bins of metallicity. Results are
shown in  Figure~\ref{fig:met}. We used the metallicities derived by \citet{Skowron2016}
for 13,016 RRab stars with a VMC counterpart and  for  4742 RRc stars we used the
metallicities derived by our LMC metallicity map.
No particular  structures are seen  in Figure~\ref{fig:met}, except for the extreme cases of metal-rich
([Fe/H]$> -0.5$ dex) and metal-poor ([Fe/H] $< -2.1$ dex) RRLs which appear to be uniformly distributed across the 
field of view, and which may belong to the MW. Like \citet{debandsingh2014} we do not find any metallicity gradient or particular
substructures related to large differences in metallicity.  

\begin{figure}
\begin{center}
\includegraphics[width=8.9cm, height=12cm]{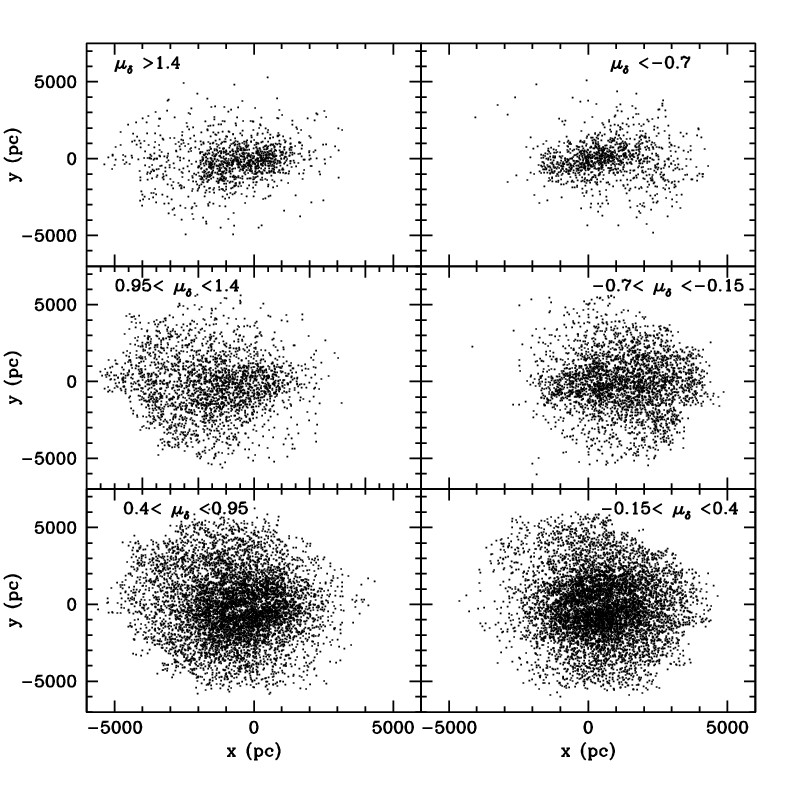}
\end{center}
\caption{LMC RRLs divided into bins of proper motion in   declination, as indicated in the labels.
}
\label{fig:pmdec}
\end{figure}

The same procedure was  applied to construct a scan map using the proper motions of the LMC RRLs available in {\it Gaia} eDR3. 
As described in Sect.~\ref{gaia}, the {\it Gaia} eDR3 catalogue includes  proper motions for 21,801 RRLs in our VMC catalogue,  
with average uncertainties of $\sigma_{\mu_{\alpha}^{*}} \sim 0.30$ mas yr$^{-1}$ and  $\sigma_{\mu_{\delta}} \sim 0.32$ mas yr$^{-1}$.
Mean proper motion values for the LMC RRLs  are 
$\mu_{\alpha}^{*}$ = $1.87\pm0.54$ mas yr$^{-1}$ and $\mu_{\delta}$ = $0.36\pm0.80$ mas yr$^{-1}$, in R.A. and Dec, respectively. 
The RRLs scan map  we constructed around these  values is shown in Figure~\ref{fig:pmdec}. 
Going clockwise from the top right-hand panel the RRLs are mapped by  
increasing the values in proper motion in the declination coordinate. 
A gradient of $\mu_{\delta}$ with  position is clearly seen. The two top  panels 
of  Figure~\ref{fig:pmdec} show the extreme cases: the spatial distribution of the RRLs is very different  with average positions of $\langle (x,y)\rangle = (-762, -152)$ pc and $\langle (x,y)\rangle = (677, -64)$ pc  for  $\mu_{\delta}$ $ > $ 1.4 mas yr$^{-1}$ and $\mu_{\delta}$ $ < $ --0.7 mas yr$^{-1}$.
This is  probably owing to the rotation of the LMC as also found by \citet{helmi2018}. 
 We performed a similar  analysis  using $\mu_{\alpha}^{*}$  obtaining a less
evident gradient, but with a detectable difference  between the south-western part  
 and  the north-eastern component of $\sim 0.25$ mas yr$^{-1}$. 

\subsection{Globular Cluster distances}
Although neither OGLE IV nor VMC were designed 
to resolve stars in clusters given their limited spatial resolution,
they detected a number of variable stars  also in the LMC star clusters.
In order to test the 3-D structure of the LMC found in  our study,   
we therefore analysed  the RRLs in 3 LMC GCs, namely, NGC~1835, NGC~2210 and NGC~1786, whose RRLs have been  observed by OGLE IV
%
(\citealt{Soszynski2016}). Those stars are mostly found in the periphery 
of the GCs in  less crowded  regions. 
These three GCs were selected because they 
are located in different regions of the LMC and observed by the VMC survey  (see 
 Figure~\ref{fig:RR_LMCmap}).
For NGC 1835 (the magenta square in Figure~\ref{fig:RR_LMCmap}),
which is located at the western edge of the LMC,
there are 9 RRLs (4 RRab and 5 RRc pulsators) in common between the OGLE IV catalogue and our VMC RRL sample.
We computed a very tight  $PW(K_\mathrm{s}$, $J-K_\mathrm{s}$) relation from these RRLs, with 
a scatter of only 0.10 mag. This confirms that the stars are all
at the same distance  with
similar  reddening and metallicity, as expected for a  GC. 
The distance to NGC~1835 we found using the RRL  $PW(K_\mathrm{s}$, $J-K_\mathrm{s}$) relation  is $d_{\rm NGC1835}$=$52.4\pm2.3$ kpc.
Three of the  four RRab stars also have  an estimate of the metallicity from  \citet{Skowron2016}.  
Their average value is   [Fe/H]$\sim -$1.84 dex, in agreement with literature values \citep[see e.g.][]{walker1993, olsen1998}. 
We applied the same procedure to NGC~2210 (the blue circle in 
 Figure~\ref{fig:RR_LMCmap}) which is located at the eastern edge of the VMC footprint. 
The $PW(K_\mathrm{s}$, $J-K_\mathrm{s}$) relation, based on 8 RRLs in the cluster, is also very tight  with 
a scatter of 0.09 mag which leads to a distance to NGC~2210 of  $d_{\rm NGC2210}$=  $47.6\pm2.3$ kpc.
Located $\sim$ 1.5 degrees to the north of NGC~1835 the GC  NGC~1786 (the green triangle in  Figure~\ref{fig:RR_LMCmap}) has 5 RRLs in common between the VMC and OGLE catalogues 
which leads to a $PW$ relation with a scatter of only 0.08 mag.
The average distance to these stars is $d_{\rm NGC1786}$=$49.9\pm2.4$ kpc.
 Although the distances of the three clusters are consistent within one sigma
there is a hint  that, as found in the previous section,   
  the eastern part of the LMC is closer to us than the western part.

\section{Discussion and Conclusions}

We have obtained new NIR  ($K_\mathrm{s}$-, $J$- and $Y$-band) and optical ($I$-band) $PL$, $PW$, 
$PLZ$ and  $PWZ$ 
relations from a sample  of $\sim 22,000$ RRLs
in the LMC using the VMC,  OGLE~IV and {\it Gaia}  data sets. 
 Our starting sample comprised more than $\sim 29,000$ RRLs with a counterpart in the VMC catalogue. However, about $\sim 7,000$ RRLs located in the central, very crowded regions of the LMC
were discarded 
using a 3-$\sigma$ clipping procedure and the  matching radius criterion, since they were overluminous with respect to the $PL/PW$ relations. A similar effect
was found in the OGLE IV data by \citet{Jacyszyn2017},  who started with an original 
sample of  27,500 RRab  stars and ended up with their  final fit being based on a subset of about  $\sim 19,500$ of them.
Our sample   of $\sim 22,000$ RRLs
 is the largest sample of RRLs 
 used to date to obtain NIR $PL$s.
 Previous NIR-PL relations available  in the literature
were based on limited samples \citep[see e.g.][ based on $\sim$40 RRLs]{Bor09}
and small fields of view  \citep[e.g.][]{sollima2006}. 
Moreover, this is the first set of $PLZ$ and $PWZ$ relations in the  $Y$-band  for RRLs in 
the LMC. 
Future facilities like James Webb Space Telescope (JWST) and The Extremely Large Telescope
(ELT) will allow the detection of RRLs in  galaxies in the Virgo cluster $\sim 20$ Mpc away from us. 
The rms of the relations we have obtained is good enough that RRLs with just one phase point on their NIR light curves will be sufficient to get an estimate of distance moduli accurate up to  $\sim 0.15$ mag.
Along with future observations 
of RRLs in galaxies outside the Local Group, our relations can thus be used to improve the calibration of the 
cosmic distance ladder.
  
 The comparison of our $PLZ$ relations  with the theoretical relations of
  \citet{marconi2015} and  with observed $PLZ$ in the literature \citep{Mur18a, neeley2019, Braga2018} indicates that the dependence on metallicity is more significant for  high metallicities. However,  the LMC lacks such metal-rich RRLs,  hence biasing the comparison  between theoretical and empirical relations.
  
 Applying the relation of \citet{Pie02} 
we derived individual reddening estimates for the  RRab stars and obtained a reddening map for the whole LMC.  This map was used to provide a reddening estimate for all RRLs in our sample.

R.A. and Dec. coordinates  of the LMC centre were derived from the average position of the RRLs and found to be at $\alpha_0 = 80^\circ.6147 ,\delta_0=-69^\circ.5799$. This 
agrees within 2$^\circ$ of almost
all centre estimates  available in the literature \citep[e.g.][based on CCs; van der Marel \& Kallivayalil 2014, for other tracers]{nikoalev2003}.
The same authors also showed that the choice of the centre influences the 
3D-shape reconstruction of the  LMC, but that  differences are rather 
small and change the determination of the  inclination and position angle by only a few degrees. 

We used our $PL_{K_{\mathrm{s}}}$  relation to estimate  individual distances to 
 $\sim22,000$ RRLs and from them we reconstructed the 3-D structure of the LMC,  as traced by these Population II stars.
The LMC resembles a regular ellipsoid with an axes ratio of 1:1.2:1.8,
where the north-eastern part of the ellipsoid is closer to us. 
The position angle and orientation of the bar identified  by \citet{Youssoufi2019} (see  Figure~5 in their paper), resembles the band devoid of RRLs seen  in our  Figure~\ref{fig:slice} ($50<d<53$ kpc panel). 
A lack of RRLs in this region is likely caused by the significant crowding conditions in the bar, which  prevents the determination of good centroids for the RRLs as well as a proper estimate of their magnitudes. 
Therefore, we did not directly detect the LMC bar  using the RRLs, in agreement with a previous study based  on OGLE~IV data and optical $PW$ relations  \citep{Jacyszyn2017}. 

A protrusion of stars extending on both sides from the centre of the LMC
might be the result of blended sources  which are still present in our sample notwithstanding the selection 
steps we performed. These blended sources are very difficult to eliminate since they mix together in every parameter space. However, it is very unlikely that this structure 
represents a  real streaming of RRLs caused by  the SMC/LMC interaction. \citet{Besla2012} presented  two models
of possible past interactions between the SMC and LMC, of which one is a direct collision. 
These models allow a good  reproduction of different peculiar 
features of both galaxies, but in neither simulations  there is a hint  of a  stellar component
projected  from the centre of the LMC along the  $|z|$ direction.

Using  metallicities from \citet{Skowron2016}  for a subsample of $\sim$ 13,000 LMC RRab stars   and for $\sim$ 5,000 RRc stars using  metal abundances derived from our metallicity map
we determined  $PLZ$ and $PWZ$ relations in the  
VMC ($Y$, $J$ and $K_\mathrm{s}$)  and optical ($I$) OGLE~IV passbands.  We also studied
the 3-D spatial distribution of the RRLs based on their  metallicities 
and we did not detect any metallicity gradient or substructure, confirming previous findings    \citep[e.g.][]{debandsingh2014}. 

{\it Gaia} eDR3 proper motions are available for a sample of $\sim 21,000$ RRLs in the VMC  catalogue. 
We used the proper motions to select RRLs which are bona fide members of the LMC and derived a 
 $PL_{K_{\mathrm{s}}}$ which is fully consistent with the relation derived from the full RRL sample.
Using the {\it Gaia} eDR3 proper motions we also  detected the rotation of the  LMC  as traced by its $>$ 10 Gyr old stars, 
in full agreement with  \citet{helmi2018} and  \citet{luri2020}  which used a larger sample ($\sim 8$ million) of  LMC stars. 
Some of the RRLs that possess large proper motions, compared with the average value, and are located on the side closer to us, can 
 also be attributed to  tidal stream components pointing towards the MW.

\begin{table*}
	\centering
	\caption{Parameters of the $PL$ and $PLZ$ relations 
	in different passbands. The coefficients are for
	the fits in the form: $m_{X0} = a + b\times\log P  + c\times$[Fe/H]}
	\label{tab:pl}
	\begin{tabular}{l  c  c  c  c c c  c  c  r } 
		\hline
		\hline
		Mode & Band &  $a$      & $\sigma_a$ & $b$         & $\sigma_b$ &   $c$   &  $\sigma_c$ & rms   & $N_{\rm star}$ \\
		\hline
		FU &  $K_{\mathrm{s0}}$  & 17.360 & 0.005    &  $-$2.84   &   0.02  & & & 0.14  &  16897   \\
		FO &  $K_{\mathrm{s}0}$  & 16.860 & 0.023     &  $-$2.98   &   0.05  & & & 0.17  &  5236    \\
      Glob &  $K_{\mathrm{s}0}$  & 17.430 & 0.004     &  $-$2.53   &   0.01  & & & 0.15  &  22122        \\
               \hline
		FU & $K_{\mathrm{s}0}$  &  17.547 &   0.008   &   $-$2.80  &   0.02     &  0.114 & 0.004   & 0.13 & 13081  \\
        FO & $K_{\mathrm{s}0}$  &  16.850 & 0.026     &  $-$2.99  &   0.05    &  0.011 & 0.012   & 0.17 & 5132  \\
	 Glob & $K_{\mathrm{s}0}$  &  17.603  & 0.007     &  $-$2.41  &   0.01    &  0.096 & 0.004   & 0.15 & 17698    \\
		       \hline
        FU &  $J_0$     & 17.699 & 0.005     &  $-$2.50    &   0.02  & & & 0.15  &  16868    \\
		FO &  $J_0$     & 17.241 & 0.022     &  $-$2.53    &   0.04  & & & 0.17  &   5225  \\
              Glob &  $J_0$     & 17.800  & 0.004     &  $-$2.00   &   0.01  & & & 0.16  &  22087  \\
                \hline
        FU &  $J_0$     &  17.888 & 0.008    & $-$2.45  &   0.02  & 0.121 &   0.004  &  0.14 & 13081  \\
        FO &  $J_0$     & 17.225 & 0.030     &  $-$2.53  &    0.05 & -0.010 &   0.012  &  0.17 & 4946 \\ 
        Glob &  $J_0$     & 17.962 & 0.007     &  $-$1.91  &    0.01 & 0.095 &   0.004  &  0.15 & 17757 \\ 
               \hline        
        FU &  $Y_0$     & 17.990 & 0.006     &  $-$2.25   &   0.02  & & & 0.17  & 16866   \\
		FO &  $Y_0$     & 17.550 & 0.025     &  $-$2.26   &   0.05  & & & 0.20  & 5211   \\
    Glob &  $Y_0$     & 18.110 & 0.046     &  $-$1.66  &   0.02  & & & 0.18  & 22071  \\  
                \hline
     FU &  $Y_0$     &  18.207 & 0.009  &  $-$2.18  &   0.03   & 0.135 & 0.005  & 0.16 & 13081  \\
         FO &  $Y_0$     &  17.553 & 0.033  &  $-$2.27  &   0.05   & 0.003 & 0.014  & 0.19 & 4946  \\
   Glob &  $Y_0$     &  18.297 & 0.008  &  $-$1.55  &   0.02   & 0.108 & 0.004  & 0.17 & 17783  \\
               \hline
                FU &  $I_0$     & 18.220  & 0.005     &  $-$2.06   &   0.02 & & & 0.13  & 16696    \\
		FO &  $I_0$     & 17.830 & 0.023     &  $-$1.99   &   0.05 & & & 0.18  & 5142      \\
              Glob &  $I_0$     & 18.350 & 0.004     &  $-$1.42   &   0.01  & & & 0.15  & 21838  \\
                \hline
              FU &  $I_0$   & 18.439  & 0.006  &   $-$1.98  &   0.02   & 0.136  & 0.003  &  0.11 & 13081 \\  
         FO &  $I_0$   & 18.439  & 0.006  &   $-$1.98  &   0.02   & 0.136  & 0.003  &  0.11 & 13081 \\  
       Glob &  $I_0$   & 18.521  & 0.006  &   $-$1.36  &   0.01   & 0.106  & 0.003  &  0.13 & 17757 \\  
            
            
                \hline
	\end{tabular}
\end{table*}

\begin{table*}
	\centering
	\caption{Parameters of the $PW$ and $PWZ$ relations in different passbands. The parameters are for 
	the fits given in the form:  $m_X - R\times(m_Y - m_X)= a+ b\times\log P  + c\times$[Fe/H] }
	\label{tab:pw}
	\begin{tabular}{l c c c c c c c c c r c} 
		\hline
		\hline
		Mode & Band & $R$ & $a$      & $\sigma_a$ & $b$       & $\sigma_b$ &  $c$ & $\sigma_c$   & rms   & $N_{\rm star}$ \\
		\hline
		FU &  $K_\mathrm{s}$, $J-K_\mathrm{s}$ & 0.69 & 17.150  &    0.006   & $-$3.075  &    0.025& &   & 0.17 & 16911 \\
		FO &  $K_\mathrm{s}$, $J-K_\mathrm{s}$ & 0.69 &  16.600    &  0.055      &  $-$3.289 &  0.027  & &   & 0.21 & 5274  \\
      Glob &  $K_\mathrm{s}$, $J-K_\mathrm{s}$ & 0.69 &  17.190 &   0.004   & $-$2.888  &   0.016 & &   & 0.18 & 22185 \\
                \hline
    FU &  $K_\mathrm{s}$, J-$K_\mathrm{s}$ & 0.69 &17.330 & 0.009     &   $-$3.033 &  0.027 &  0.111 &  0.004 &  0.16 & 13081   \\
     FO &  $K_\mathrm{s}$, J-$K_\mathrm{s}$ & 0.69 & 16.619   & 0.033     &  $-$3.280 &  0.052 &  0.008 &  0.013 &  0.19 & 4945   \\
      Glob&  $K_\mathrm{s}$, J-$K_\mathrm{s}$ & 0.69 & 17.348   & 0.008     &  $-$2.810 &  0.016 &  0.094 &  0.004 &  0.17 & 17670   \\
         \hline 
        FU   &  $K_\mathrm{s}$, $Y-K_\mathrm{s}$ & 0.42 &   17.110   &    0.006   &   $-$3.086   &  0.024   & & & 0.16 & 16911    \\
		FO   &  $K_\mathrm{s}$, $Y-K_\mathrm{s}$ & 0.42 & 16.580  &   0.026    &  $-$3.276   &   0.053 & & & 0.20 & 5274  \\
		Glob &  $K_\mathrm{s}$, $Y-K_\mathrm{s}$ & 0.42 & 17.150  & 0.004     & $-$2.888   &   0.015 & & & 0.17 & 22185 \\
                     \hline
	FU &  $K_\mathrm{s}$, $Y-K_\mathrm{s}$ & 0.42 & 17.298  & 0.009 &  $-$3.049  &   0.026  &  0.107 & 0.004 & 0.15 & 13081  \\
	FO &  $K_\mathrm{s}$, $Y-K_\mathrm{s}$ & 0.42 & 16.591  & 0.031      & $-$3.270  &  0.050    &  0.008 & 0.013 & 0.18 & 4946  \\	
    Glob &  $K_\mathrm{s}$, $Y-K_\mathrm{s}$ & 0.42 & 17.306  & 0.007      & $-$2.813  &  0.016    &  0.088 & 0.004 & 0.16 & 17757  \\	
                       \hline      
        FU &  $K_\mathrm{s}$, I-$K_\mathrm{s}$ & 0.25 &  17.160 &  0.006    &  $-$3.036  &   0.023  & &  & 0.16 & 16686   \\
		FO &  $K_\mathrm{s}$, I-$K_\mathrm{s}$ & 0.25 &  16.610  &  0.026     &  $-$3.243  &    0.052 & &   & 0.20 &  5142    \\
      Glob &  $K_\mathrm{s}$, I-$K_\mathrm{s}$ & 0.25 & 17.210   &  0.004   &  $-$2.800  &    0.015  & &   & 0.17 & 21543   \\  
                    \hline
		FU &  $K_\mathrm{s}$, I-$K_\mathrm{s}$ & 0.25 &  17.337  &  0.009  & $-$2.999 &  0.027&  0.110  & 0.004 &  0.15 &  13081    \\
		FO &  $K_\mathrm{s}$, I-$K_\mathrm{s}$ & 0.25 &  16.629 &  0.033     &  $-$3.217 &  0.052 &  0.004  & 0.013 &  0.19 &  4937    \\
    	Glob &  $K_\mathrm{s}$, I-$K_\mathrm{s}$ & 0.25 &  17.368 &  0.008     &  $-$2.711 &  0.016 &  0.091  & 0.004 &  0.16 &  17567    \\
                     \hline    
        FU &  $K_\mathrm{s}$, $V-K_\mathrm{s}$ & 0.13 &  17.170    &   0.005   &  $-$3.021 &  0.022 & &   & 0.15 & 16686    \\
		FO &  $K_\mathrm{s}$, $V-K_\mathrm{s}$ & 0.13 &  16.630  &    0.024   &  $-$3.216   &   0.048  & &   & 0.18 & 5142     \\
        Glob &  $K_\mathrm{s}$, $V-K_\mathrm{s}$ & 0.13 & 17.220   &  0.004   &  $-$2.787   &   0.014  & &  & 0.16 &  21828  \\
                     \hline
         FU &  $K_\mathrm{s}$, $V-K_\mathrm{s}$ & 0.13 &  17.349  &  0.008     & $-$2.985 &  0.025 & 0.109   &  0.004 &  0.14  & 13081  \\
         FO &  $K_\mathrm{s}$, $V-K_\mathrm{s}$ & 0.13 &  16.650 &  0.031     & $-$3.193  &  0.049 & 0.004   &  0.013 &  0.18  & 4957  \\
        Glob &  $K_\mathrm{s}$, $V-K_\mathrm{s}$ & 0.13 &  17.380 &  0.007      & $-$2.701  &  0.015 & 0.090   &  0.004 &  0.15  & 17230  \\     
                    \hline      
        FU &  $J$, $I-J$ & 0.96 &   17.190   &     0.008    &  $-$2.921   &   0.035 & &   &  0.23 & 16686  \\
		FO &  $J$, $I-J$ & 0.96 &    16.660       &    0.033   &  $-$3.056   &  0.066 & &   &  0.25 & 5142    \\
        Glob &  $J$, $I-J$ & 0.96 &      17.270    &  0.006   & $-$2.549   &  0.022  & &   &  0.24 & 21828  \\
                     \hline      
		FU &  $J$, $I-J$ & 0.96 &     17.358  &  0.014   &  $-$2.903  &   0.041 &  0.106 &  0.007 &  0.23   & 13081 \\
		FO &  $J$, $I-J$ & 0.96 &    16.658  & 0.043     &  $-$3.040 &   0.068 &  -0.007 &  0.017 &  0.25   & 13081 \\
		Glob &  $J$, $I-J$ & 0.96 &         17.425  & 0.011     &  $-$2.431 &   0.023 &  0.085 &  0.006 &  0.23   & 17757 \\	
                      \hline
        FU &  $J$, $V-J$ & 0.40 &       17.220    & 0.006    &   $-$2.920  &   0.027& &    &  0.18 & 16686 & \\
		FO &  $J$, $V-J$ & 0.40 &      16.700    &   0.025     & $-$3.038  &  0.051  & &    &  0.20 &  5142  &  \\
        Glob &  $J$, $V-J$ & 0.40 &      17.280   &  0.005    &  $-$2.593  &  0.017 & &    &  0.18 & 21828 & \\
                     \hline     
         FU &  $J$, $V-J$ & 0.40 &  17.377 &   0.010   &   $-$2.897 &  0.031  &  0.103 &  0.005 &  0.17 & 13081    \\
		FO &  $J$, $V-J$ & 0.40 & 16.697  &  0.033     &  $-$3.027 &  0.052 &  $-$0.007 &  0.013 &  0.20 & 4946    \\
		FU &  $J$, $V-J$ & 0.40 & 17.433  &  0.009     &  $-$2.493 &  0.017 &  0.084 &  0.005 &  0.18 & 17546    \\	
                  \hline           
                    
        FU &  $I$, $V-I$ & 1.55 &    17.160 &   0.004   &   $-$3.016  &   0.018     & & & 0.12 & 16686  \\
		FO &  $I$, $V-I$ & 1.55 &   16.690   &    0.018     &     $-$3.101  &    0.037   & & & 0.14  & 5142  \\
        Glob &  $I$, $V-I$ & 1.55 &   17.190 &    0.003  & $-$2.843  &   0.011     & & &0.13 &  21828\\
                     \hline
        FU &  $I$, $V-I$ & 1.55 & 17.299 &  0.006  &  $-$2.994  &  0.019   & 0.090 & 0.003 & 0.11 & 13081  \\
        FO &  $I$, $V-I$ & 1.55 & 16.712  &  0.024     & $-$3.105    &  0.037    & 0.018 & 0.010 & 0.11& 4954  \\
        Glob &  $I$, $V-I$ & 1.55 & 17.323  &  0.006     & $-$2.790   &  0.012    & 0.076 & 0.003 & 0.12 & 17456  \\
		\hline
		\hline
	\end{tabular}
\end{table*}

\section*{Acknowledgments}
We thank the Cambridge Astronomy Survey Unit (CASU) and the Wide Field Astronomy
Unit (WFAU) in Edinburgh for providing calibrated data products under  support of the
Science and Technology Facility Council (STFC) in the UK. 
M-RC acknowledges support from the European Research Council (ERC) under European Union’s Horizon 202 research and innovation programme (grant agreement no. 682115).
This work makes use of data 
from the ESA mission {\it Gaia} (https://www.cosmos.esa.int/gaia), processed by
the {\it Gaia} Data Processing and Analysis Consortium (DPAC, https://www.cosmos.esa.int/web/gaia/dpac/consortium). 
Funding for the DPAC has been provided by national institutions, in particular
the institutions participating in the {\it Gaia} Multilateral Agreement.

\section*{Data availability}

The data underlying this article are available in the article and in its online supplementary material or will be shared on reasonable request to the corresponding author.

\label{lastpage}

\end{document}